\documentclass[aps,pra,showpacs,amsmath,amssymb,superscriptaddress,longbibliography,reprint,10pt]{revtex4-2}
\usepackage{graphicx}
\usepackage{xcolor}
\usepackage{newtxtext,newtxmath,nccmath}
\usepackage{mathrsfs}
\usepackage{braket}
\usepackage[colorlinks=true,breaklinks=true,allcolors=blue]{hyperref}
\usepackage{bm}

\definecolor{blue}{HTML}{008ED7}
\definecolor{mygray}{gray}{0.75}
\definecolor{lightBlue2}{rgb}{0.94, 0.97, 1.0}
\definecolor{lightBlue}{HTML}{8fa6d2}
\definecolor{darkBlue}{HTML}{243d9a}

\newcommand{\appsection}[1]{\section{\MakeUppercase{#1}}}

\allowdisplaybreaks

\usepackage{orcidlink}

\definecolor{emma}{rgb}{0,0,1}

\definecolor{giovanna}{rgb}{1,0,1}

\definecolor{red}{rgb}{1,0,0}

\begin{document}
\title{Optimal spatial searches with long-range tunneling}

\author{Emma C. King \orcidlink{0000-0002-6696-3235}}
\affiliation{Theoretische  Physik,  Universit\"at  des  Saarlandes,  D-66123  Saarbr\"ucken,  Germany}

\author{Moritz Linnebacher \orcidlink{0009-0002-5544-4461}}
\affiliation{Theoretische  Physik,  Universit\"at  des  Saarlandes,  D-66123  Saarbr\"ucken,  Germany}

\author{Peter P. Orth \orcidlink{0000-0003-2183-8120}}
\affiliation{Theoretische  Physik,  Universit\"at  des  Saarlandes,  D-66123  Saarbr\"ucken,  Germany}

\author{Matteo Rizzi \orcidlink{0000-0002-8283-1005}}
\affiliation{Institut f\"ur Theoretische Physik, Universit\"at zu K\"oln, D-50937 K\"oln, Germany}

\author{Giovanna Morigi \orcidlink{0000-0002-1946-3684}}
\affiliation{Theoretische  Physik,  Universit\"at  des  Saarlandes,  D-66123  Saarbr\"ucken,  Germany}

\date{\today}

\begin{abstract}
A quantum walk on a lattice is a paradigm of a quantum search in a database. The database qubit strings are the lattice sites, qubit rotations are tunneling events, and the target site is tagged by an energy shift. For quantum walks on a continuous time, the walker diffuses across the lattice and the search ends when it localizes at the target site. The search time $T$ can exhibit Grover's optimal scaling with the lattice size $N$, namely, $T\sim \sqrt{N}$, on an all-connected, complete lattice. For finite-range tunneling between sites, instead, Grover's optimal scaling is warranted when the lattice is a hypercube of $d>4$ dimensions. Here, we show that Grover's optimum can be reached in lower dimensions on lattices of long-range interacting particles, when the interaction strength scales algebraically with the distance $r$ as $1/r^{\alpha}$ and $0<\alpha<3d/2$. For $\alpha<d$ the dynamics mimics the one of a globally connected graph. For $d<\alpha<d+2$, the quantum search on the graph can be mapped to a short-range model on a hypercube with spatial dimension $d_s=2d/(\alpha-d)$, indicating that the search is optimal for $d_s>4$. Our work identifies an exact relation between criticality of long-range and short-range systems, it provides a quantitative demonstration of the resources that long-range interactions provide for quantum technologies, and indicates when existing experimental platforms can implement efficient analog quantum search algorithms.
\end{abstract}

\maketitle

\section{Introduction}

Grover's quantum search algorithm is a paradigm of quantum computing advantage as it holds the promise of a quadratic speedup with respect to a classical search~\cite{Grover1996_FastQuantumMechanical,Grover1997_QuantumMechanicsHelps,Zalka1999_GroverQuantumSearching}. In fact, for an unstructured search the computational time scales with the number of entries $N$ of a database as $T_Q\sim \sqrt{N}$, which shall be compared with the classical counterpart, where $T_C\sim N$. The quantum algorithm thus ideally belongs to a more favorable time complexity class. The latter refers to the scaling of the time needed to run an algorithm with the size of the database, and is a central concept of complexity theory in computer science. Practical overheads of digital quantum circuits can severely reduce the advantage with respect to the classical counterpart \cite{Preskill:2018}. In this context, analog quantum platforms---such as those leveraging continuous-time quantum walks---may exhibit reduced overhead and inherent resilience to certain types of noise, 
allowing one to fully exploit the speedup provided by quantum mechanics. 

The analog implementation of Grover's quantum search in continuous time realizes a quantum walk on a complete graph, whose vertices are the elements of the database and in which the target state is tagged by an energy shift \cite{Farhi_Gutmann1998_AnalogAnalogueDigital}.
Starting from a uniform superposition of all the graph's vertices, the dynamics localizes the walker in the target site over a computational time $T\sim \sqrt{N}$. Global connectivity is a sufficient but not necessary condition: Grover's optimal scaling can be achieved in strongly regular graphs \cite{Meyer_Wong2015_ConnectivityPoorIndicator, Janmark_etal2014_GlobalSymmetryUnnecessary,Cattaneo:2018,Wong_etal2018_QuantumWalkSearch}, random graphs satisfying certain requirements on the edges' probability distribution \cite{Chakraborty_etal2016_SpatialSearchQuantum,Chakraborty_etal2017_OptimalQuantumSpatial}, and in spatial searches on hypercubes with dimension \mbox{$d>4$} and nearest-neighbor coupling between vertices \cite{Childs_Goldstone2004_SpatialSearchQuantum,Aaronson_Ambainis2003_QuantumSearchSpatial}. In the latter case, hypercubes with $d<4$ fail to realize the Grover speedup.

\begin{figure}[b]
    \centering
    \includegraphics[width=\linewidth]{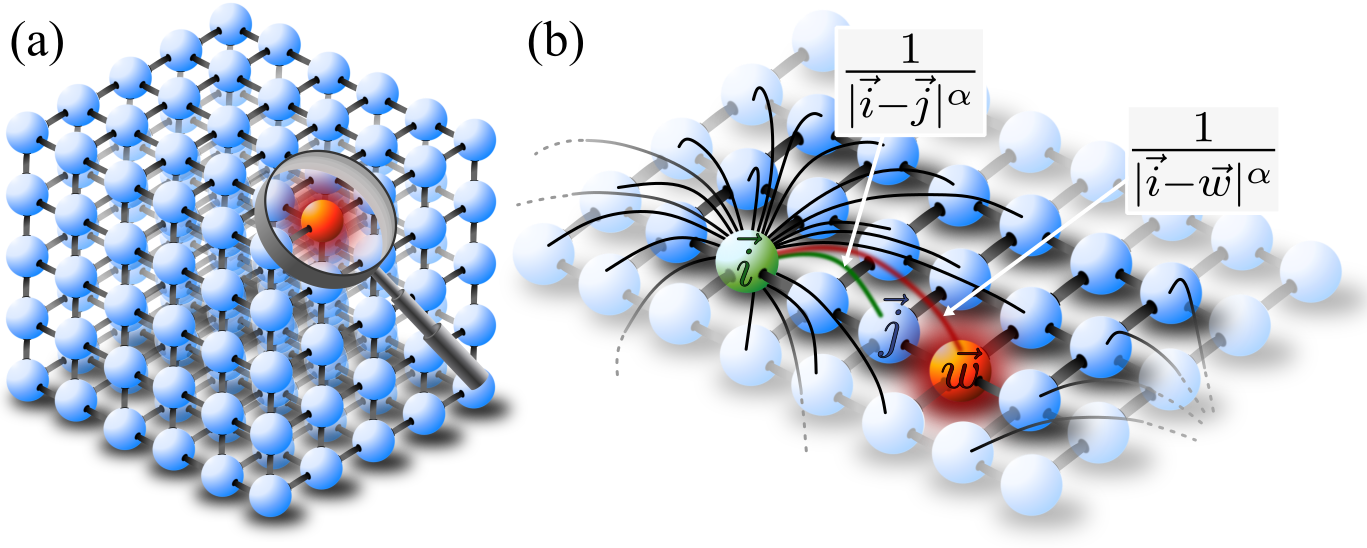}
    \caption{(a) Illustrative graphic of search on a cubic lattice (hypercube with $d=3$) with nearest-neighbor couplings ($\alpha\rightarrow\infty$). Target node is depicted in red.  (b) Schematic of the power-law scaling of the connectivity of a single site $\vec{i}$ in a two-dimensional cubic lattice.
    }
    \label{fig:schematic}
\end{figure}

Spatial searches on hypercubes establish a direct connection between quantum search algorithms and dynamical localization on lattices, implemented in photonic architectures \cite{Schreiber_etal2011_DecoherenceDisorderQuantum,Nitsche:2016,Graefe:2020} and dipole traps \cite{Karski:2009,Preiss_etal2015_StronglyCorrelatedQuantum,Young:2022}. The requirement of $d>4$ spatial dimensions, on the other hand, is a demanding condition. Yet, most physical implementations are based on particles whose interactions are longer-ranged and scale with the distance $r$ as $1/r^\alpha$ with $\alpha>0$ \cite{Defenu_etal2023_LongrangeInteractingQuantum}. These systems can realize long-range $XY$ models, where a single excitation performs a quantum walk with the tunneling amplitude scaling as $1/r^\alpha$ \cite{Tran_etal2021_OptimalStateTransfer,Lewis_etal2021_OptimalQuantumSpatial}. The resulting graph is all-connected, whereby the edge capacities decrease with the Euclidean distance between the connected vertices, as illustrated in Fig.~\ref{fig:schematic}. In abstract terms, the exponent $\alpha$ of the edges' algebraically-decaying capacity interpolates between two very distinct limits of the time complexity of the spatial search: the nearest-neighbor case ($\alpha\to\infty$), where Grover speedup is found for hypercubes of dimension  $d>4$, and the complete graph ($\alpha=0$), where the notion of spatial dimension is lost. Clarifying under which conditions long-range interacting systems are a resource for quantum algorithms, and what the intimate connection between graph connectivity and dimensionality is, would permit to assess the computational resources of these systems and to identify advantageous experimental implementations.

In this work we show that simple arrays of long-range interacting atoms or molecules, in which interactions scale algebraically with the distance $r$ as $1/r^\alpha$, can implement an optimal analog quantum search \cite{Anikeeva_etal2021_NumberPartitioningGrover}. We identify the condition on the exponent $\alpha$ and on the spatial dimension $d$, for which Grover optimal scaling can be realized. This is done by performing a formal mapping between a long-range quantum walk on a cubic lattice of dimension $d\le 3$ and a short-range quantum walk on a lattice in spatial dimension $d_s=d_s(d,\alpha)$, corresponding to the spectral dimension of the graph. In doing so, we also provide a formal demonstration of the role of the spectral dimension in connecting criticality of long-range systems with the corresponding short-range ones for a dynamics that is central to several quantum algorithms. 

\section{Search by quantum walk in continuous time} 

The database comprises $N$ sites on a $d$-dimensional hypercube, with vertices (nodes) identified in space by the $d$-dimensional vector $\vec{i}\equiv (i_1,\ldots,i_d)$ with 
$i_j=1,\ldots,n$ and $n=N^{1/d}$. The states $\{|i_1,\ldots,i_d\rangle\}$ form a basis of the Hilbert space of the single walker. Denoting by $|w\rangle$ the target state, the dynamics is governed by the dimensionless Hamiltonian 
\begin{equation}\label{eq:H}
\hat H_{\alpha}=-\gamma_0 L_\alpha-|w\rangle\langle w|\,, 
\end{equation}
with the Laplacian $L_\alpha$ encoding the graph's properties. In the formulation of Ref.\ \cite{Farhi_Gutmann1998_AnalogAnalogueDigital}, the graph is complete. Correspondingly, the Laplacian is $L_0= N\vert s\rangle\langle s\vert$, where $|s\rangle=\sum_{i_1,\ldots ,i_n}|\vec{i}\rangle/\sqrt{N}$ is the uniform superposition of all states of the database. Preparing the database in the extended state $|s\rangle$ reduces the dynamics to a two-dimensional Hilbert space, consisting of the state $|r\rangle = \sum_{i_1,\ldots ,i_n\neq w}|\vec{i}\rangle/\sqrt{N-1}$ and the localized state $|w\rangle$. In the reduced Hilbert space the eigenvalues of $\hat{H}_\alpha$ have an energy gap $\Delta E=\sqrt{4 \gamma_0(1-N) +(\gamma_0  N + 1)^2}$ and the transfer amplitude at time $t$, $A(t)=\langle w|{\rm e}^{-{\rm i}\hat{H}_\alpha t}|s\rangle$, is a rotation in the two-dimensional subspace. We denote by $T$ the minimal time at which the fidelity $F(T)=|A(T)|^2$ reaches unity. This time scales as $T\propto1/\Delta E=\sqrt{N}/2$ when $\gamma_0$ is set at the critical point $\gamma_c=1/N$ separating the extended ground state $|s\rangle$ from the localized ground state $|w\rangle$. Next, we address the open question of the conditions under which such a dramatic reduction in the dynamically explored Hilbert space dimension occurs in long-range tunneling models of arbitrary dimension $d$.

\subsection{Power-law tunneling} 

Assume now that the Laplacian of Eq.\ \eqref{eq:H} has the form 
\begin{equation}\label{eq:L}
L_\alpha=\sum_{\vec{i}\neq\vec{j}}\frac{1}{|\vec{i}-\vec{j}|^\alpha}\left(|\vec{i}\rangle\langle\vec{j}|+{\rm H.c.}\right)-\varepsilon_0 I\,,
\end{equation}
with $\vert\dots\vert$ the Euclidean norm of the vector, $\alpha>0$ the tunneling power-law exponent, and $\varepsilon_0$ a real scalar, multiplying the identity $I$ and to be determined. With this Laplacian, Hamiltonian \eqref{eq:H} generalizes the dynamics to all-connected graphs but with edge capacities that decay algebraically with the Euclidean distance, as illustrated in Fig.~\ref{fig:schematic}(b). This family of graphs includes the complete graph in the limit $\alpha\to 0$. In the opposite limit, $\alpha\to\infty$, the corresponding graph is a hypercube with nearest-neighbor coupling in $d$ spatial dimensions: Grover's optimum is reached for $d>4$ after setting $\varepsilon_0=0$ and choosing $\gamma_0$ at the critical point of the phase transition between the extended ground state $|s\rangle$ and the localized state $|w\rangle$ \cite{Childs_Goldstone2004_SpatialSearchQuantum}. 

We now determine the time complexity class as a function of $\alpha$ and $N$, namely, the scaling of the time $T$ with $N$ and $\alpha$, where $T$ now is the time at which the transfer probability amplitude $A(T)$ reaches its first maximum \footnote{We do not consider the scaling with the error $\epsilon$ with which the fidelity may deviate from unity. See Ref.\ \cite{Chakraborty_etal2020_OptimalitySpatialSearch} for a discussion.}. For this purpose, we  first generalize
the derivation of Ref.\ \cite{Childs_Goldstone2004_SpatialSearchQuantum} to power-law Laplacians~\eqref{eq:L}. This is achieved by decomposing the transfer amplitude $A(t)$ using the eigenstates $|\psi_i\rangle$ of $\hat{H}_\alpha$ at eigenvalue $E_i$:
\begin{equation}
\label{Eq:A(T)}
A(t)=\sum_{i=1}^{N}\mathcal W_i^{\phantom{*}}\mathcal S_i^* {\rm e}^{-iE_it}\approx \sum_{i=0,1}\mathcal W_i^{\phantom{*}}\mathcal S_i^* {\rm e}^{-iE_it},
\end{equation}
where ${\mathcal W_i}=\langle w|\psi_i\rangle$ and ${\mathcal S_i}=\langle s|\psi_i\rangle$ are the overlaps of the eigenstate $|\psi_i\rangle$ with the target and extended state, respectively. We now identify the conditions for which the sum in Eq.~\eqref{Eq:A(T)} can be reduced to the sole contribution of ground and first excited states.

\subsection{Conditions for reduction in Hilbert space}\label{app:A1}

\begin{figure}[t]
    \centering
    \includegraphics[width=1\columnwidth]{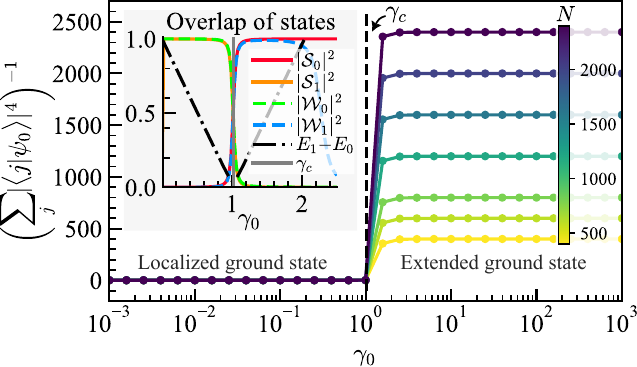}
    \caption{Participation ratio $1/\sum_j\vert\langle j \vert\psi_0\rangle \vert^4$ for the ground state $\vert\psi_0\rangle$ \textit{vs} $\gamma_0$, with the critical value $\gamma_0=\gamma_c$ separating the extended from the localized ground state. Inset shows scalar products $\vert\mathcal{S}_{i=0,1}\vert^2$ and $\vert\mathcal{W}_{i=0,1}\vert^2$ and the energy gap $E_1-E_0$ as a function of parameter $\gamma_0$. We set $d=1$ and $\alpha=0.6$; qualitatively similar results hold for other dimensions and tunneling exponents.}
    \label{fig:A1}
\end{figure}

Information about the scalar products ${\mathcal W_i}$ and $\mathcal S_i$ in Eq.~\eqref{Eq:A(T)} is extracted from the projections $\langle j|\hat{H}_\alpha|\psi_i\rangle =E_i\langle j|\psi_i\rangle$. It is convenient to make use of the transcendental equation $\mathcal{F}_\alpha(E_i) = 1$, whose roots are the eigenvalues $E_i$ \cite{Childs_Goldstone2004_SpatialSearchQuantum}. The transcendental equation is derived by recasting the time-independent Schr\"odinger equation for the search Hamiltonian $\hat{H}_\alpha$ as $$\vert \psi_i\rangle = {\mathcal W_i}(-\gamma_0 L_\alpha - E_i)^{-1}\vert w\rangle,$$ and taking the projection onto $|w\rangle$ \footnote{It can be shown that $(-\gamma_0 L_\alpha - E_i)$ is invertible, and that states with $\langle w \vert\psi_i\rangle=0$ can be safely neglected since they do not 
contribute to the fidelity.}. It reads
\begin{equation}\label{eq:F}
		\mathcal{F}_\alpha(E) = \frac{1}{N} \sum_{\vec{k}\in \mathrm{BZ}} \frac{1}{\gamma_0 \mathcal{E}_\alpha(\vec{k})-E}\,,
\end{equation}
where $\mathcal{E}_\alpha(\vec{k}) =\sum_{\vec j\neq \vec{0}} \cos(\vec{k}\cdot\vec{j})/|\vec j|^\alpha -\varepsilon_0 $ are the eigenvalues of the Laplacian $L_\alpha$ and the sum runs over the first Brillouin zone. The expression for the overlap $\mathcal{W}_i$ is found from the normalization condition $\langle\psi_i|\psi_i\rangle=1$, giving $|\mathcal{W}_i|^2=1/\mathcal{F}'(E_i)$, with $\mathcal{F}_\alpha'(E_i)\equiv d\mathcal{F}_\alpha(E)/dE|_{E=E_i}$. Moreover, $|\mathcal S_i|^2=|\mathcal W_i|^2/N E_i^2$, and the probability amplitude takes the compact form 
\begin{equation}
A(t) = \frac{1}{\sqrt{N}} \sum_i \frac{1}{\vert E_i\vert \mathcal{F}_\alpha'(E_i)} e^{-i E_i t}\,.
\end{equation}
The behavior of the transcendental function permits to identify the condition when the sum can be reduced to the first two terms. We first assume that $|E_0|,|E_1|\ll \gamma_0\mathcal E_{\alpha}(\vec{k})$ for all \mbox{$\vec{k}\neq 0$}, namely, that there is an energy gap between the two lowest eigenvalues and the rest of the spectrum. This requires setting $\varepsilon_0=\sum_{\vec j\neq \vec{0}} 1/|\vec j|^\alpha-\mathcal{E}_{\alpha}(\vec{0})$. In order to find an expression for the overlaps, we perform the Taylor expansion of $\mathcal F(E)$ and of its derivative:
\begin{equation}\label{eq:F_expansion}
		\mathcal{F}_\alpha(E_{i}) \approx \frac{-1}{N E_i} + \frac{S_{1}^{(\alpha)}}{\gamma_0} + \frac{S_{2}^{(\alpha)}}{\gamma_0^2}E_i\,, \quad \mathcal{F}_\alpha'(E_{i})\approx \frac{1}{N E_i^2} + \frac{S_{2}^{(\alpha)}}{\gamma_0^2}\,,
\end{equation}
with $i=0,1$ and $S_\ell^{(\alpha)} = \sum_{\vec{k}\neq\vec{0}} [\mathcal{E}_\alpha(\vec{k})]^{-\ell}/N$, Eq.~\eqref{eq:S}. 
Using that the 
overlap $|\mathcal{W}_i|^2=1/\mathcal{F}'(E_i)$, then $\mathcal W_0\approx \mathcal W_1$ and $\mathcal S_0\approx \mathcal S_1$. Together with the truncated expression for $\mathcal{F}_\alpha'(E_{i=0,1})$, this provides an equation connecting the two lowest energy eigenvalues: 
\begin{equation}
-E_0\left(\frac{1}{NE_0^2}+\frac{S_2^{(\alpha)}}{\gamma_0^2}\right)+ E_1\left(\frac{1}{NE_1^2}+\frac{S_2^{(\alpha)}}{\gamma_0^2}\right)\approx 0\,.
\end{equation}
Solving for $\gamma_0$ gives $\gamma_0 = \sqrt{E_0E_1 N S_2^{(\alpha)}}$. To make the expression explicit, one must insert the energies, which themselves depend on the parameter $\gamma_0$. We extract the energies from the truncated form of transcendental equation $\mathcal{F}(E)=1$ in Eq.~\eqref{eq:F_expansion}: \mbox{$E_i = \gamma_0/\left(2S_2^{(\alpha)}\right) \left[\gamma_0 -S_1^{(\alpha)} \pm \sqrt{ (S_1^{(\alpha)}-\gamma_0 )^2+4 S_2^{(\alpha)}/N}\right]$}. Inserting this into the expression for $\gamma_0$ gives the trivial solution $\gamma_0=0$ and the critical value $\gamma_0=\gamma_c=S_1^{(\alpha)}$, separating the extended from the localized ground state; see Fig.~\ref{fig:A1}. 

This analysis shows that the sum in Eq.~\eqref{Eq:A(T)} can be reduced to the sole contribution of ground and first excited states, $|\psi_0\rangle$ and $|\psi_1\rangle$, respectively, after setting $\varepsilon_0=\sum_{\vec j\neq 0} 1/|\vec j|^\alpha-\mathcal{E}_\alpha(\vec{0})$ and $\gamma_0$ to the value $\gamma_c=S_1^{(\alpha)}$. Here, 
\begin{equation}\label{eq:S}
S_\ell^{(\alpha)} = \sum_{\vec{k}\neq\vec{0}} [\mathcal{E}_\alpha(\vec{k})]^{-\ell}/N\,,    
\end{equation}
where the sum runs over the first Brillouin zone of the hypercube and $\mathcal{E}_\alpha(\vec{k})$ are the eigenvalues of the Laplacian $L_\alpha$ \eqref{eq:L}. For $\gamma_0=\gamma_c$ the two lowest energies  read $E_{1}=-E_0=\chi_\alpha/\sqrt{N}$, where 
\begin{equation}\label{eq:chi}
    \chi_\alpha=S_1^{(\alpha)}/\sqrt{S_2^{(\alpha)}}\,,
\end{equation}
with $|\chi_\alpha|\le 1$. The transfer amplitude is approximated by the expression $A(t)\approx \chi_\alpha \sin (\chi_\alpha t/\sqrt{N})$ and the first maximum is reached at $T\approx \pi\sqrt{N}/(2 \chi_\alpha)$ with fidelity $F(T)=|\chi_\alpha|^2$. Grover's optimum is reached when $\chi_\alpha\to 1$.

\subsection{Time complexity and spectral gap} 

The scaling $T\sim \sqrt{N}$ is obtained when the two lowest energies are well separated from the rest of the spectrum. For the choice of parameters in which the ground state of the Laplacian is almost degenerate with the target state, this condition can be reformulated as $E_1-E_0\ll \delta_\alpha$, where $\delta_\alpha=\mathcal E_\alpha(\vec k_1)-\mathcal E_\alpha(\vec{0})$ is the spectral gap of the Laplacian, namely, the energy difference between the Laplacian's ground state and the first excited eigenstate at $|\vec{k}_1|=2\pi/N^{1/d}$. The spectral gap is proportional to the \textit{algebraic connectivity} $a(G)=\mathcal{E}_\alpha(\vec{k}_1)$ of the graph $G$ \cite{Fiedler73,expander}, whose magnitude provides indications on the graph's robustness. Increases of the algebraic connectivity have been linked to shorter characteristic path lengths between sites, resulting in faster propagation \cite{Chung1997,Millan_etal2021_LocalTopologicalMoves}. Interestingly, the condition for the optimal search sets a lower bound on the spectral gap: 
\begin{equation}\label{eq:gap_condition}
    \delta_\alpha>c_\alpha/\sqrt{N}\,,
\end{equation}
where $c_\alpha$ is a positive constant and we have used that $E_1-E_0=2\chi_\alpha/\sqrt{N}$ at the optimal value $\gamma_0=\gamma_c$. We note that the same condition has been found for the optimal search in a random graph \cite{Chakraborty_etal2020_OptimalitySpatialSearch}, suggesting that this is the necessary requirement that a generic graph shall satisfy for achieving Grover's optimum. 

We now derive the asymptotic scaling of the spectral gap. In order to provide a scaling behavior with a well-defined thermodynamic limit, we analyze the rescaled gap $\Delta_\alpha=\delta_\alpha/\mathcal{E}_\alpha(\vec{k}_{\max})$ with $\vert \vec{k}_{\max}\vert = \sqrt{d}\pi$. This rescaling corresponds to the operation $L_\alpha\to L_\alpha'=L_\alpha/\mathcal{E}_\alpha(\vec{k}_{\max})$ and $\gamma_0\to \gamma_0'=\gamma_0\mathcal{E}_\alpha(\vec{k}_{\max})$, which does not change the Hamiltonian but now makes the rescaled bandwidth and critical value independent of $N$. In the rescaled framework, $\gamma_c'=\gamma_c\mathcal{E}_\alpha(\vec{k}_{\max})$ is finite for $N\to\infty$ and is the critical point of the quantum phase transition separating the extended from the localized ground state. We restrict our analysis to spatial dimensions $d\le4$ and find that
\begin{equation}\label{eq:gap_scaling}
    \Delta_\alpha \approx
    \begin{cases}
        1-\mathscr{C}^{(d)}_{1}(\alpha) & \alpha\in[0,d)\\
        \mathscr{C}^{(d)}_{2}(\alpha)N^{1-\alpha/d}& \alpha\in(d,d+2)\\
        \mathscr{C}^{(d)}_{3}(\alpha)N^{-2/d}& \alpha\ge d+2
    \end{cases}
\end{equation}
with $\alpha$-dependent constants $\mathscr{C}^{(d)}_{i=1,2,3}(\alpha)$ of order $1$ and accurate up to a factor $\mathcal{C}\in [1,d^{\alpha/2}]$. The detailed derivation, including the explicit functional behavior of the coefficients, is relegated to Appendixes~\ref{s:spectral_gap_scaling} and \ref{ss:detailed_derivations}. Three distinct cases emerge. {\it (i)} For $\alpha < d$, the gap $\Delta_\alpha$ is independent of $N$. Therefore, condition \eqref{eq:gap_condition} is satisfied for sufficiently large $N$. Using other words, for $\alpha<d$ the power-law tunneling becomes sufficiently long-ranged to mimic a globally connected graph. {\it (ii)} For $\alpha\in (d,d+2)$ the spectral gap scales as $N^{1-\alpha/d}$ and the condition on optimality depends on the value that $\alpha$ takes within the interval $(d,d+2)$. Finally, 
{\it (iii)} for \mbox{$\alpha>d+2$} the spectral gap scales as $N^{-2/d}$ and the quadratic speedup is provably lost. This includes the nearest-neighbor tunneling of Ref.~\cite{Childs_Goldstone2004_SpatialSearchQuantum} in the limiting case $\alpha\to \infty$.  

Case {\it (ii)} is peculiar, since it identifies an additional exponent $\alpha_c=3d/2$ separating the behavior of the gap: For $d+2>\alpha>\alpha_c$ the quantity $\delta_\alpha\sqrt{N}$ decreases with $N$, while for $\alpha_c>\alpha>d$ it exhibits the opposite behavior (cf.\ Eq.\ \eqref{eq:gap_scaling}). Hence, at fixed dimension $d\le 4$, when $\alpha < \alpha_c$, condition~\eqref{eq:gap_condition} holds asymptotically, and the Grover optimum is attainable. For $\alpha > \alpha_c$, instead, the gap condition is violated, and the search time $T$ reverts to the suboptimal scaling. 

\begin{figure}[t]
    \centering
    \includegraphics[width=\linewidth]{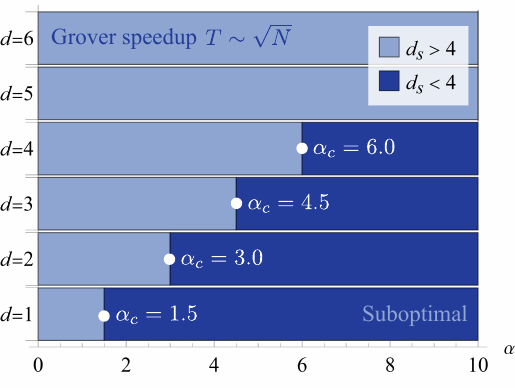}
    \caption{Search optimality diagram illustrating the connection between the spatial dimension $d\in\mathbb{Z}$, in which the hypercube is embedded, and the power-law tunneling exponent $\alpha$, and how it relates to the time complexity. The critical exponent $\alpha_c=3d/2$, represented by the white points, separates regimes in which we have optimal ($d_s>4$, \textcolor{lightBlue}{light blue}) and suboptimal ($d_s<4$, \textcolor{darkBlue}{dark blue}) quantum spatial search, distinguished by the critical spectral dimension $d_{s}=4$.}
    \label{fig:phase_diagram}
\end{figure}
\begin{figure*}[t]
	\includegraphics[width=0.9\textwidth]{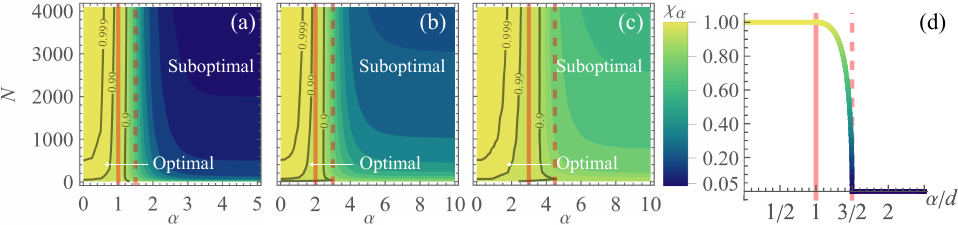}
	\caption{Upper bound to the search fidelity: $\chi_\alpha$ as a function of the number of lattice sites $N$ and long-range tunneling exponent $\alpha$ for (a) $d=1$, (b) $d=2$, and (c) $d=3$. Vertical lines correspond to $\alpha=d$ (solid) and $\alpha=\alpha_c=3d/2$ (dashed). Contours are shown for $\chi_\alpha=0.999,\,0.99,\,0.9$. (d) Asymptotic behavior of $\chi_\alpha$, Eq.\ \eqref{eq:chi}, in the limit $N\rightarrow\infty$ as a function of $\alpha/d$. In Appendix \ref{app:fidelity} we show that, for $0<\alpha<d$, $\chi_\alpha=1$, while for $d<\alpha<3d/2$ it decreases monotonically to zero as $\chi_\alpha=\sqrt{3-2\alpha/d}/(2-\alpha/d)$.}
	\label{fig:chi}
\end{figure*}

Equation \eqref{eq:gap_scaling} permits to establish an insightful connection between $\alpha$ and $d$ in determining the time complexity of the quantum search, which we summarize in Fig.~\ref{fig:phase_diagram}. This behavior is linked to the Lieb-Robinson bound, namely, with the maximal speed at which information can propagate across the lattice \cite{Deshpande_etal2018_DynamicalPhaseTransitions,Lieb_Robinson72,Tran_etal2020_HierarchyLinearLight}. It proves that the critical exponent $\alpha_c=3d/2$ is associated with a phase transition in the time complexity of the search problem. The order parameter is $\chi_\alpha$, Eq.\ \eqref{eq:chi}, its squared value is the fidelity, that quantifies the occupation of the two lower energy eigenstates and is reminiscent of the order parameter of Bose-Einstein condensation in interacting systems \cite{Penrose:1956}. The dependence of $\chi_\alpha$ on $\alpha$ and $d$ is reported in Fig.\ \ref{fig:chi}(a)--(c) for a range of values of $N$. The asymptotic limit, for $N\to\infty$, is displayed in Fig.\ \ref{fig:chi}(d) as a function of $\alpha/d$. The value $\alpha_c=3d/2$ separates the regime at $\alpha>\alpha_c$, where the search dynamics spans over a size of the Hilbert space with dimension of the order $N$, from the condensed regime at $\alpha<\alpha_c$, where the search occurs in the subspace consisting of the extended state $|s\rangle$ and the target state $|w\rangle$. As such, at $\alpha=\alpha_c$ a phase transition occurs, separating regimes with different time complexities. This result applies to hypercubes with spatial dimension $d\le 4$ and arbitrary sizes. It thus also encompasses the specific limit $d=1$, where it rigorously proves and generalizes conclusions drawn in Ref.\ \cite{Lewis_etal2021_OptimalQuantumSpatial} for finite one-dimensional chains. 

\section{Spectral dimension and criticality} \label{s:spectral_dimension_criticality}

We now focus on the regime where $\alpha\in (d,d+2)$ and perform a formal mapping of the long-range model of Eq.\ \eqref{eq:H} to an \textit{effective} short-range model with a higher dimension $D$ that exhibits the same critical behavior. The existence of this kind of mapping has been conjectured in the context of criticality of long-range systems \cite{Angelini_etal2014_RelationsBetweenSRLRIsing,Defenu_etal2023_LongrangeInteractingQuantum, Bighin_etal2024_UniversalScalingReal}. We now prove it for the extended-localized quantum phase transition of Hamiltonian \eqref{eq:H}. In order to proceed, it is convenient to introduce the concept of \textit{spectral dimension} $d_s\in\mathbb{R}^+$ \cite{Alexander_Orbach1982_DensityStatesFractals,Burioni_Cassi1996_UniversalPropertiesSpectral}, which characterizes a fictitious diffusion process on the search space. This quantity can be interpreted as the dimension \textit{perceived by} the quantum walker and, as we now demonstrate, coincides with the effective dimension $D$ of an equivalent model that only includes nearest-neighbor couplings.

The spectral dimension $d_s$ can be extracted when the density of states scales as a power-law  in the low-energy limit. Denoting by $\lambda=\mathcal{E}_\alpha(\vec{k})/\mathcal E_{\alpha}(\vec{k}_{\rm max})$ the rescaled eigenvalues, then for $\lambda\ll1$ the density of states scales as $\rho(\lambda) \sim \lambda^{\beta}$. The associated cumulative distribution $\rho_{\rm{CD}}(\lambda)$ according to the relation is $$\rho_{\rm{CD}}(\lambda)=\int_0^\lambda \rm{d}\Lambda\, \rho(\Lambda) \sim \lambda^{d_s/2}\,,$$
which defines the spectral dimension $d_s=2(\beta+1)$. This connects the finite-size scaling of the gap with the spectral dimension: In fact, for a gap $\Delta$ that vanishes in the thermodynamic limit, $\rho_{\rm{CD}}(\Delta)=1/N\sim \Delta^{d_s/2}$. Assume now a nearest-neighbor model. In this case, the gap vanishes as $N^{-2/D}$ and $\rho_{\rm{CD}}(\Delta)\sim \Delta^{D/2}$. This establishes a direct link between $D$ and $d_s$, and thus between the universal critical behavior of the short-range and of the long-range model. In the regime $\alpha\in(d,d+2)$, by means of the spectral gap scaling \eqref{eq:gap_scaling} we can extract the spectral dimension: 
	\begin{equation}\label{eq:ds}
		d_s = \frac{2d}{\alpha-d} = \frac{2d}{\sigma}\,, %
	\end{equation}
with $\sigma=\alpha-d$. For $\alpha\in (d,d+2)$, the spectral dimension \mbox{$d_s=4$} is the upper critical dimension of the long-range quantum walk. This identifies the critical exponent $\alpha=\alpha_c=3d/2$, such that for $\alpha>\alpha_c$ at $\gamma_c$ the ground state ceases to transition to a truly localized ground state.  

Equation \eqref{eq:ds} provides a simple means to assess the time complexity of the quantum walk for lower-dimensional lattices, $d\le 4$. For algebraically-decaying interactions with exponent $\alpha\in (d,d+2)$, the long-range model in $d$ dimensions can be mapped to a short-range model in $D=d_s$ dimensions. One can then use the predictions of Ref.\ \cite{Childs_Ge2014_SpatialSearchContinuoustime} valid for continuous walks on a hypercube with short-range hopping, and Grover's optimal search is achieved for $d_s>4$. We note that in the limit $\alpha\rightarrow d^+$ the spectral dimension \eqref{eq:ds} diverges: In this limit, in fact, the power-law tunneling becomes sufficiently long-ranged to mimic a globally connected graph with $\alpha=0$. In contrast, at $\alpha= d+2$ we obtain $d_s=d$, and the effective dimension $D$ reduces to the real spatial dimension~$d$.

\section{Experimental implementations}

The relations of Sec.~\ref{s:spectral_dimension_criticality} have practical implications on identifying physical platforms for analog quantum searches. In fact, $XY$ Hamiltonians are simulated by arrays of Rydberg atoms \cite{Morgado:2021,Bluvstein_etal2022_QuantumProcessorBased}. Spin models with long-range coupling coefficients are being realized in chains of trapped ions \cite{Richerme_etal14,Jurcevic_etal14,Monroe_etal2021_ProgrammableQuantumSimulations}, as well as in ultracold atoms in optical cavities \cite{Periwal_etal2021_ProgrammableInteractionsEmergent}. Individual atom addressing permits to `tag' a target qubit and to perform local measurements \cite{Anikeeva_etal2021_NumberPartitioningGrover,Monroe_etal2021_ProgrammableQuantumSimulations,Bluvstein:2023,tao2024high}. Our work demonstrates that these platforms can realize an efficient analog quantum search provided that the spatial dimensionality of the lattice satisfies the relation $\alpha<3d/2\le 6$. For instance, Rydberg interactions ($\alpha=6$) \cite{Morgado:2021,Bluvstein:2023} might realize an efficient analog search provided that the lattice geometry has the same connectivity as a hypercube in four dimensions.  Figure \ref{fig:chi}, moreover, suggests that a quantum advantage can be reached for databases of finite size. For dipolar interactions Grover's time complexity can be reached in a simple-cubic lattice in three dimensions \cite{Baier_etal2016_ExtendedBoseHubbardModels,Chomaz:2023}, while any scalable Coulomb crystalline structure, from three dimensions down to the Coulomb chain \cite{Dubin_ONeil_1999_RevModPhys_TrappedNonneutralPlasmas,Bohnet:2016}, can warrant Grover's optimum. In light of the fact that in certain platforms the interaction exponent is tunable \cite{Monroe_etal2021_ProgrammableQuantumSimulations,Periwal_etal2021_ProgrammableInteractionsEmergent}, one could perform a constrained optimization of exponent and lattice dimensionality \eqref{eq:ds} taking into account the experimental overhead.

\section{Outlook}  

In this work, we have determined the time complexity of quantum searches on a lattice with long-range tunneling. We have considered tunneling amplitudes scaling with the distance $r$ as $1/r^\alpha$ and identified the relation between exponent $\alpha$ and dimensionality $d$ for which Grover optimal scaling is achieved, $\alpha\le 3d/2$. We have focused here on lattices with cubic geometries, however, we anticipate the bound $d_s=4$ to apply more generally to other search spaces. The results could also be further extended to search on fractals, where it has been conjectured that the spectral dimension, as opposed to the fractal dimension, is the appropriate metric to quantify the search's optimality \cite{Agliari_etal2010_QuantumwalkApproachSearching,Sato_etal2020_ScalingHypothesisSpatial,Patel_Raghunathan2012_SearchFractalLattice}. 

From a broader perspective, this study brings to the fore the tight connection between criticality and quantum search algorithms. A key concept is the spectral dimension $d_s$: The value $d_s=4$ is the upper critical dimension of the localized-extended quantum phase transition and of the dynamical phase transition to the quadratic scaling of Grover's optimum. The critical exponent $\alpha_c=3d/2$ separates an ergodic to a localized phase. Since the spectral dimension $d_s$ has been conjectured to govern universality in non-homogeneous systems \cite{Cassi_1992,Wu_1995,Bighin_etal2024_UniversalScalingReal}, this study can additionally offer valuable insights into search optimality on diverse structures. One interesting question is whether similar critical behavior appears in other complexity measures, such as Solomonoff Kolmogorov Chaitin (SKC) complexity, which quantifies the shortest algorithmic description of a problem \cite{Solomonoff1964,Kolmogorov1968,Chaitin1969}. A highly compressible search space with structured spectral properties (low SKC complexity) may correspond to more efficient quantum evolution and faster search times.

These concepts could be key for other kinds of dynamics based on quantum searches, such as optimization problems like, for instance, active learning agents \cite{Paparo_etal2014_QuantumSpeedupActive}.
Our insights allows to quantitatively assess the resources that long-range interactions offer for quantum search algorithms, and might provide a key to determine the resource they can offer for quantum information processing and quantum technologies in general~\cite{Defenu_etal2023_LongrangeInteractingQuantum,Deshpande_etal2018_DynamicalPhaseTransitions}.


\section*{Acknowledgments} E.C.K.\ and G.M.\ acknowledge helpful discussions with Markus Bl\"{a}ser on the algebraic connectivity of graphs and on time complexity of quantum algorithms. This work was funded by the Deutsche Forschungsgemeinschaft (DFG, German Research Foundation)---Project-ID 429529648-TRR306 QuCoLiMa (``Quantum Cooperativity of Light and Matter'') and the QuantERA project ``QNet: Quantum transport, metastability, and neuromorphic applications in Quantum Networks''---Project ID 532771420, by the German Ministry of Education and Research (BMBF) via the Project NiQ (Noise in Quantum Algorithms), and the
Deutsche Forschungsgemeinschaft (DFG, German Research Foundation) under Germany’s Excellence Strategy---EXC 2004/1-390534769---Cluster of Excellence ML4Q (``Matter and Light for Quantum Computing''). This research was supported in part by the National Science Foundation Grants No.\ NSF PHY-1748958 and No. NSF PHY-2309135 to the Kavli Institute for Theoretical Physics (KITP).

\section*{Data availability} Some of the data that support the findings of this article are openly available \cite{Code}. All other data is such that it is
interoperable and reusable.

\appendix
\numberwithin{equation}{section}
\numberwithin{figure}{section}

\appsection{Asymptotic scaling of the Laplacian spectral gap}\label{s:spectral_gap_scaling}
The asymptotic scaling of the spectral gap $\Delta_\alpha=\big[\mathcal{E}_\alpha(\vec{k}_1) - \mathcal{E}_\alpha(\vec{0})\big]/\mathcal{E}_\alpha(\vec{k}_{\max})$ of the rescaled Laplacian is reported in Eq.~\eqref{eq:gap_scaling}. Here we provide analytic expressions for the prefactors $\mathscr{C}^{(d)}_{i=1,2}(\alpha)$ for $d=1,\dots,4$, and compare the results with exact numeric data. First, note that $\mathcal{E}_\alpha(\vec{k}) =\sum_{\vec j\neq \vec{0}} \cos(\vec{k}\cdot\vec{j})/|\vec j|^\alpha -\varepsilon_0 $ is the eigenenergy of the Laplacian $L_\alpha$ in the unscaled framework, with $\varepsilon_0=\sum_{\vec j\neq 0} 1/|\vec j|^\alpha-\mathcal{E}_\alpha(\vec{0})$ a constant energy shift. The momentum vector $\vec{k}_{\max}$ is defined by its magnitude $\vert \vec{k}_{\max}\vert = \sqrt{d}\pi$. It follows that
\begin{equation}\label{eq:SMgap}
    \Delta_\alpha = \frac{\delta_\alpha}{\mathcal{E}_\alpha(\vec{k}_{\max})} = \frac{\sum_{\vec j\neq 0} \cos(\vec{k}_1\cdot\vec{j})/|\vec j|^\alpha -\kappa_0 }{\mathcal{E}_\alpha(\vec{k}_{\max})}\,,
\end{equation}
where $\delta_\alpha\equiv\mathcal{E}_\alpha(\vec{k}_1)-\mathcal{E}_\alpha(\vec{0})$ and $\kappa_0\equiv\varepsilon_0+\mathcal{E}_\alpha(\vec{0})$. The scaling behavior of $\Delta_\alpha$ is determined by analyzing the numerator and denominator of Eq.~\eqref{eq:SMgap} independently. To make the calculations analytically tractable, we describe the distance between the hypercube graph vertices using the Manhattan norm, a $p$-norm $\vert\vert\vec{x}\vert\vert_p = \big(\sum_{i=1}^d \vert x_i\vert^p \big)^{1/p}$ with $p=1$. This gives the length of the shortest path constrained to move along the graph's edges (i.e.\ a ``grid distance''), compared to the Euclidean norm $\vert\vec{x}\vert \equiv \vert\vert\vec{x}\vert\vert_2$ which was used in the main text and describes the shortest distance between the two vertices. Since all norms in finite-dimensional spaces are equivalent, we can make a statement about the scaling and prefactor of the spectral gap independent of the norm that is chosen. To make this more concrete, using the Cauchy-Schwarz and H\"older's inequality, we find the relation between $p$-norms as $\vert\vert\vec{x}\vert\vert_p\leq \vert\vert\vec{x}\vert\vert_r \leq d^{1/r-1/p}\vert\vert\vec{x}\vert\vert_p$, $0<r<p$. Therefore, $\vert \vec{j}\vert \leq \vert\vert\vec{j}\vert\vert_1 \leq \sqrt{d} \vert \vec{j}\vert$, and the scaling of the spectral gap \eqref{eq:SMgap} is evidently unaffected by the choice of norm. We proceed by using the Manhattan norm. Constants $\mathscr{C}^{(d)}_{i=1,2}(\alpha)$ reported in Table~\ref{tab:spectral_gap_rescaled} are accurate up to a factor $\mathcal{C}\in [1,d^{\alpha/2}]$ when implementing the Euclidean norm.

In Table \ref{tab:spectral_gap_unscaled} we summarize the large-$N$ asymptotic results for $\delta_\alpha$ and $\mathcal{E}_\alpha(\vec{k}_{\max})$ of Eq.~\eqref{eq:SMgap}, with detailed derivations and numeric checks provided in Sec.~\ref{ss:detailed_derivations}. With these results, we extract the scaling behavior with $N$ of the spectral gap. Two regimes need to be considered: $\alpha<d$ and $d<\alpha<d+2$. For $\alpha<d$, both the numerator and denominator of Eq.~\eqref{eq:SMgap} diverge in the limit $N\rightarrow\infty$. Applying l'Hôpital's rule, we evaluate the $N\rightarrow\infty$ limit. Previously of an indeterminate form, the asymptotic limit of the spectral gap is converted into a limit that can be evaluated directly. This leads to explicit analytic expressions for the constants $\mathscr{C}^{(d)}_1(\alpha)$ of the main text, all of which are independent of $N$, see Eqs.~\eqref{eq:spectral_gap_D1_asymptotic_1}, \eqref{eq:spectral_gap_D2_asymptotic_1}, \eqref{eq:spectral_gap_D3_asymptotic_1} and \eqref{eq:spectral_gap_D4_asymptotic_1} of Table~\ref{tab:spectral_gap_rescaled}. 

To extract the scaling in the regime $d<\alpha<d+2$ we require a different approach. Note that for $\alpha>d$ the term scaling as $\sim N^{1-\alpha/d}$ in $\kappa_0$ asymptotically satisfies $c N^{1-\alpha/d}\ll 1$, with $c$ an $\alpha$-dependent prefactor, as reported in Table \ref{tab:spectral_gap_unscaled} (see Eqs.~\eqref{eq:kappa0_D1_asymptotic}, \eqref{eq:kappa0_D2_asymptotic}, \eqref{eq:kappa0_D3_asymptotic}, and \eqref{eq:kappa0_D4_asymptotic}).
 Defining $x\equiv c N^{1-\alpha/d}$, we expand the spectral gap~\eqref{eq:SMgap} around $x=0$. Generically, the spectral gap~\eqref{eq:SMgap} written in terms of $x$ takes the form $\Delta_\alpha\approx \frac{x+f(\alpha,N)}{x+g(\alpha)}$, with $f(\alpha,N)\sim N^{1-\alpha/d}$ and $g(\alpha)$ simply an $\alpha$-dependent constant. Performing now the small-$x$ expansion, we obtain
\begin{equation}
	\Delta_\alpha \approx \frac{f}{g} + \frac{(g-f)}{g^2} x + \mathcal{O}(x^2)\,,
\end{equation} 
where the first term $f/g$ is the leading-order contribution. From Table \ref{tab:spectral_gap_unscaled} we insert the appropriate functions $f(\alpha,N)$ and $g(\alpha)$, leading to the asymptotic results for the spectral gap, $\Delta_\alpha\approx f/g$. These results give the analytic expressions for the prefactors $\mathscr{C}^{(d)}_2(\alpha)$ of the main text, see Eqs.~\eqref{eq:spectral_gap_D1_asymptotic_2}, \eqref{eq:spectral_gap_D2_asymptotic_2}, \eqref{eq:spectral_gap_D3_asymptotic_2} and \eqref{eq:spectral_gap_D4_asymptotic_2} of Table~\ref{tab:spectral_gap_rescaled}.

\begin{center}
\begin{table*}
\renewcommand{\arraystretch}{1.2}
\caption{Asymptotic scaling of the (unscaled) spectral gap $\delta_\alpha$, the largest eigenenergy $\mathcal{E}_\alpha(\vec{k}_{\max})$ for normalization, and the energy shift $\kappa_0\equiv\varepsilon_0+\mathcal{E}_\alpha(\vec{0})$. Results are reported for $d\in[1,4]$ and $\alpha<d+2$ to leading-order in $N$, with sub-dominant contributions of order $\mathcal{O}(N^{1-1/d-\alpha/d})$ being neglected. For compactness, we introduced the following $N$-independent function: $\mathcal{K}_{n}(z) \equiv E_{n}(i\pi z) + E_{n}(-i\pi z)$, where $E_n(x)$ is the exponential integral function. Note that $\zeta(s)$, $\Gamma(s)$ and $_pF_q(a;b;z)$ denote the Riemann $\zeta$ function, $\Gamma$ function and generalized hypergeometric function, respectively. For $\alpha>d+2$ the well-established asymptotic scaling of the gap $\delta_\alpha\sim N^{-2/d}$ is recovered and is not reported in this table. }%
\label{tab:spectral_gap_unscaled}
\begin{tabular}{p{0.16\textwidth} p{0.84\textwidth}}
\hline
\hline
Dimension $d=1$ &  \multicolumn{1}{c}{$\alpha<3$} \\
\hline
Unscaled spectral gap: &\vspace{-0.3cm}
\begin{equation}\label{eq:spectral_gap_D1_asymptotic_unscaled}
\delta_\alpha \approx -\kappa_0 + 2\zeta(\alpha)
+2^\alpha \pi^{\alpha-1}\sin\!\left(\frac{\pi \alpha}{2}\right)\Gamma(1-\alpha)\, N^{1-\alpha}
\end{equation}\vspace{-0.5cm}
\\
Largest eigenenergy: &\vspace{-0.3cm}
\begin{equation}\label{eq:Epi_D1_asymptotic}
\mathcal{E}_\alpha(k_{\max}) \approx -\varepsilon_0 + (2^{2-\alpha}-2)\zeta(\alpha)
\end{equation}\vspace{-0.5cm}
\\
Constant energy shift: &\vspace{-0.3cm}
\begin{equation}\label{eq:kappa0_D1_asymptotic}
\kappa_0 \approx 2\zeta(\alpha) + \frac{-2^\alpha}{\alpha-1} \, N^{1-\alpha}
\end{equation}
\\
\hline
Dimension $d=2$ & \multicolumn{1}{c}{$\alpha<4\ (\alpha\neq1)$} \\
\hline
Unscaled spectral gap: &\vspace{-0.3cm}
\begin{equation}\label{eq:spectral_gap_D2_asymptotic_unscaled}
\delta_\alpha \approx -\kappa_0 +4\zeta(\alpha-1)
-\Biggl[\frac{2^{\alpha }\, _1F_2\!\left(\frac{2-\alpha}{2};\frac{1}{2},\frac{4-\alpha}{2};-\frac{\pi ^2}{4}\right)}{(\alpha -2) (\alpha -1)}
+\frac{ 2\mathcal{K}_{\alpha -1}(2)  - 2^{\alpha-1}\mathcal{K}_{\alpha -1}(1) }{(\alpha -1)} \Biggr] N^{1-\alpha/2}
\end{equation}\vspace{-0.5cm}
\\
Largest eigenenergy: &\vspace{-0.3cm}
\begin{equation}\label{eq:Epi_D2_asymptotic}
\mathcal{E}_\alpha(\vec{k}_{\max}) \approx -\varepsilon_0 + (2^{4-\alpha}-4)\zeta(\alpha-1)
\end{equation}\vspace{-0.5cm}
\\
Constant energy shift: &\vspace{-0.3cm}
\begin{equation}\label{eq:kappa0_D2_asymptotic}
\kappa_0 \approx 4\zeta(\alpha-1) + \frac{4-2^{\alpha+1}}{(\alpha-2)(\alpha-1)} N^{1-\alpha/2}
\end{equation}
\\
\hline
Dimension $d=3$ & \multicolumn{1}{c}{$\alpha<5\ (\alpha\neq1,2)$} \\
\hline
Unscaled spectral gap: &\vspace{-0.3cm}
\begin{equation}\label{eq:spectral_gap_D3_asymptotic_unscaled}
\begin{aligned}
\delta_\alpha \approx{}& -\kappa_0 +4\zeta(\alpha-2) +2\zeta(\alpha) +\Biggl[-\frac{2^{\alpha }  \, _1F_2\!\left(\frac{3-\alpha}{2};\frac{1}{2},\frac{5-\alpha}{2};-\frac{\pi ^2}{4}\right)}{(\alpha -3) (\alpha -2) (\alpha -1)} \\
&\qquad\qquad\qquad\qquad\qquad\qquad\quad + \frac{2^{\alpha} \mathcal{K}_{\alpha -2}(1)-4 \mathcal{K}_{\alpha-2}( 2)-2^{\alpha -1} 3^{3-\alpha } \mathcal{K}_{\alpha -2}(3)}{(\alpha -2) (\alpha -1)} \Biggr] N^{1-\alpha/3}
\end{aligned}
\end{equation}\vspace{-0.5cm}
\\
Largest eigenenergy: &\vspace{-0.3cm}
\begin{equation}\label{eq:Epi_D3_asymptotic}
\mathcal{E}_\alpha(\vec{k}_{\max}) \approx -\varepsilon_0 + (2^{5-\alpha}-4)\zeta(\alpha-2)+(2^{2-\alpha}-2)\zeta(\alpha)
\end{equation}\vspace{-0.5cm}
\\
Constant energy shift: &\vspace{-0.3cm}
\begin{equation}\label{eq:kappa0_D3_asymptotic}
\kappa_0 \approx 4\zeta(\alpha-2)+2\zeta(\alpha) - \frac{3^{1-\alpha}(9(2^\alpha)-8(3^\alpha)+6^\alpha)}{(\alpha-3)(\alpha-2)(\alpha-1)} N^{1-\alpha/3}
\end{equation}
\\
\hline
Dimension $d=4$& \multicolumn{1}{c}{$\alpha<6\ (\alpha\neq1,2,3)$} \\
\hline
Unscaled spectral gap: &\vspace{-0.3cm}
\begin{equation}\label{eq:spectral_gap_D4_asymptotic_unscaled}
\begin{aligned}
\delta_\alpha \approx{}& -\kappa_0 + \frac{8}{3}\zeta (\alpha -3)+\frac{16}{3}\zeta (\alpha -1) - \Biggl[\frac{2^{\alpha } \, _1F_2\!\left(\frac{4-\alpha}{2};\frac{1}{2},\frac{6-\alpha}{2};-\frac{\pi ^2}{4}\right)}{(\alpha -4)(\alpha -3) (\alpha -2) (\alpha -1)} \\ 
&\qquad\qquad\qquad\qquad\qquad -\frac{3\cdot 2^{\alpha -1} \mathcal{K}_{\alpha-3}(1)-2^{\alpha } 3^{4-\alpha } \mathcal{K}_{\alpha-3}(3)-2^{7-\alpha }\mathcal{K}_{\alpha-3}(4)}{(\alpha -3) (\alpha -2) (\alpha -1)}
\Biggr] N^{1-\alpha/4}
\end{aligned}
\end{equation}\vspace{-0.5cm}
\\
Largest eigenenergy: &\vspace{-0.3cm}
\begin{equation}\label{eq:Epi_D4_asymptotic}
\mathcal{E}_\alpha(\vec{k}_{\max}) \approx -\varepsilon_0 + \frac{1}{3}\!\left(2^{7-\alpha}-2^3\right)\zeta(\alpha-3) +\frac{1}{3}\!\left(2^{6-\alpha}-2^4\right)\zeta(\alpha-1)
\end{equation}\vspace{-0.5cm}
\\
Constant energy shift: &\vspace{-0.3cm}
\begin{equation}\label{eq:kappa0_D4_asymptotic}
\kappa_0 \approx \frac{8}{3} \zeta (\alpha -3) +\frac{16}{3}\zeta (\alpha -1)   -\frac{4 \left(-2^{6-\alpha }+2^{\alpha }+2^{\alpha } 3^{4-\alpha }-24\right)}{(\alpha -4) (\alpha -3) (\alpha -2) (\alpha -1)}N^{1-\alpha/4}
\end{equation}
\\
\hline
\end{tabular}
\end{table*}
\end{center}

\begin{table*}
\caption{Analytic expressions for the constants $\mathscr{C}^{(d)}_{i=1,2}(\alpha)$ appearing in the asymptotic scaling of the spectral gap: $\Delta_\alpha \approx 1-\mathscr{C}^{(d)}_{1}(\alpha),\, \alpha\in[0,d)$, and $\Delta_\alpha\approx\mathscr{C}^{(d)}_{2}(\alpha)N^{1-\alpha/d},\, \alpha\in(d,d+2)$; see Eq.~\eqref{eq:gap_scaling} of the main text. For compactness, we introduced the following $N$-independent functions: $h(\alpha)=(2^{2-\alpha}-4)\zeta(\alpha)$, $f(\alpha)=(2^{4-\alpha}-8)\zeta(\alpha-1)$, $g(\alpha) = (2^{5-\alpha}-8)\zeta(\alpha-2)+(2^{2-\alpha}-4)\zeta(\alpha)$, $j(\alpha) = -3^{-1} 2^{4-\alpha } \left(\left(2^{\alpha }-8\right) \zeta (\alpha -3)+2 \left(2^{\alpha }-2\right) \zeta (\alpha -1)\right)$, and $\mathcal{K}_{n}(z) \equiv E_{n}(i\pi z) + E_{n}(-i\pi z)$, where $\zeta(s)$ is the Riemann $\zeta$ function and $E_n(x)$ is the exponential integral function. The notation $\Gamma(s)$ and $_pF_q(a;b;z)$ is used to denote the $\Gamma$ function and generalized hypergeometric function, respectively. Constants $\mathscr{C}^{(d)}_{i=1,2}(\alpha)$ are exact (asymptotically) for $d=1$, and accurate up to a factor $\mathcal{C}\in [1,d^{\alpha/2}]$ for $d=2,3,4$.}
\label{tab:spectral_gap_rescaled}
\centering
\renewcommand{\arraystretch}{1.2}
\begin{tabular}{p{0.49\textwidth} p{0.49\textwidth}}
\hline
\hline
\multicolumn{1}{l}{Dimension $d=1$\qquad $\alpha<1$} &
\multicolumn{1}{l}{Dimension $d=1$\qquad $1<\alpha<3$} \\
\hline
\begin{equation}\label{eq:spectral_gap_D1_asymptotic_1}
\mathscr{C}^{(1)}_1(\alpha) = (1-\alpha)\pi^{\alpha-1}\sin\left(\frac{\pi\alpha}{2}\right)\Gamma(1-\alpha)
\end{equation}
&
\begin{equation}\label{eq:spectral_gap_D1_asymptotic_2}
\mathscr{C}^{(1)}_2(\alpha) =
\left[\frac{2^\alpha \pi^{\alpha-1} (\alpha-1) \sin(\pi\alpha/2)\Gamma(1-\alpha)+2^\alpha}{h(\alpha)\,(\alpha-1)} \right]
\end{equation}
\\
\hline
\multicolumn{1}{l}{Dimension $d=2$\qquad $\alpha<2$} &
\multicolumn{1}{l}{Dimension $d=2$\qquad $2<\alpha<4$} \\
\hline
\begin{equation}\label{eq:spectral_gap_D2_asymptotic_1}
\begin{aligned}
\mathscr{C}^{(2)}_1(\alpha) =&\ \frac{_1F_2\!\left(\frac{2-\alpha }{2};\frac{1}{2},\frac{4-\alpha }{2};-\frac{\pi ^2}{4}\right)}{2-2^{2-\alpha}} \\
&- \frac{(\alpha-2)\mathcal{K}_{\alpha -1}(2)}{2-2^\alpha}
+\frac{(\alpha-2)\mathcal{K}_{\alpha -1}(1)}{2^{3-\alpha}-4}
\end{aligned}
\end{equation}
&
\begin{equation}\label{eq:spectral_gap_D2_asymptotic_2}
\begin{aligned}
\mathscr{C}^{(2)}_2(\alpha) = \frac{1}{f(\alpha)}\Bigg[&
\frac{2^{\alpha+1}-4}{(\alpha-2)(\alpha-1)}
- \frac{ 2\mathcal{K}_{\alpha -1}(2) -2^{\alpha-1} \mathcal{K}_{\alpha -1}(1)}{(\alpha -1)} \\
&- \frac{2^\alpha\, _1F_2\!\left(\frac{2-\alpha}{2};\frac{1}{2},\frac{4-\alpha}{2};-\frac{\pi ^2}{4}\right)}{(\alpha-2)(\alpha-1)} \Bigg]
\end{aligned}
\end{equation}
\\
\hline
\multicolumn{1}{l}{Dimension $d=3$\qquad $\alpha<3$} &
\multicolumn{1}{l}{Dimension $d=3$\qquad $3<\alpha<5$} \\
\hline
\begin{equation}\label{eq:spectral_gap_D3_asymptotic_1}
\begin{aligned}
\mathscr{C}^{(3)}_1(\alpha) =&\
\frac{ _1F_2\!\left(\frac{3-\alpha}{2};\frac{1}{2},\frac{5-\alpha}{2};-\frac{\pi ^2}{4}\right)}{3\!\left(3^{2-\alpha}-2^{3-\alpha}+1\right)} -(\alpha-3) \\
&\times\frac{2^{\alpha +1}  \mathcal{K}_{\alpha -2}(1)-8 \mathcal{K}_{\alpha -2}(2)-2^{\alpha } 3^{3-\alpha } \mathcal{K}_{\alpha -2}(3)}{6 \left(3^{2-\alpha } 2^\alpha + 2^{\alpha }-8\right)}
\end{aligned}
\end{equation}
&
\begin{equation}\label{eq:spectral_gap_D3_asymptotic_2}
\begin{aligned}
\mathscr{C}^{(3)}_2(\alpha) =&\ \frac{1}{g(\alpha)}\Bigg[
-\frac{2^{\alpha }  \, _1F_2\!\left(\frac{3-\alpha}{2};\frac{1}{2},\frac{5-\alpha}{2};-\frac{\pi ^2}{4}\right)}{(\alpha -3) (\alpha -2) (\alpha -1)} \\
&+ \frac{2^{\alpha } \mathcal{K}_{\alpha -2}(1)-4 \mathcal{K}_{\alpha -2}(2)-2^{\alpha -1} 3^{3-\alpha } \mathcal{K}_{\alpha -2}(3)}{(\alpha -2) (\alpha -1)} \\
&+ \frac{3^{1-\alpha}(9(2^\alpha)-8(3^\alpha)+6^\alpha)}{(\alpha-3)(\alpha-2)(\alpha-1)}\Bigg]
\end{aligned}
\end{equation}
\\
\hline
\multicolumn{1}{l}{Dimension $d=4$\qquad $\alpha<4$} &
\multicolumn{1}{l}{Dimension $d=4$\qquad $4<\alpha<6$} \\
\hline
\begin{equation}\label{eq:spectral_gap_D4_asymptotic_1}
\begin{aligned}
\mathscr{C}^{(4)}_1(\alpha) =&\
\frac{2^{-2}\, _1F_2\!\left(\frac{4-\alpha }{2};\frac{1}{2},\frac{6-\alpha }{2};-\frac{\pi ^2}{4}\right)}
{ \left(-3\ 2^{\alpha +3}+4^{\alpha }-64\right)4^{-\alpha} +3^{4-\alpha}} \\
& -\frac{ 3\ 4^{\alpha } (\alpha-4) \left( \mathcal{K}_{\alpha -3}(1) -54\ 3^{-\alpha } \mathcal{K}_{\alpha -3}(3)\right)}%
{8  \left( -3\ 2^{\alpha +3}+4^{\alpha }-4^3 +3^{4-\alpha}\ 4^{\alpha }\right)} \\
& +\frac{ 2^8 (\alpha-4) \mathcal{K}_{\alpha -3}(4)}%
{8  \left( -3\ 2^{\alpha +3}+4^{\alpha }-4^3 +3^{4-\alpha}\ 4^{\alpha }\right)}
\end{aligned}
\end{equation}
&
\begin{equation}\label{eq:spectral_gap_D4_asymptotic_2}
\begin{aligned}
\mathscr{C}^{(4)}_2(\alpha) =&\ \frac{1}{j(\alpha)}\Bigg[
-\frac{2^{\alpha }  \, _1F_2\!\left(\frac{4-\alpha }{2};\frac{1}{2},\frac{6-\alpha }{2};-\frac{\pi ^2}{4}\right)}{(\alpha-4)(\alpha -3) (\alpha -2) (\alpha -1)} \\
&+ \frac{3\ 2^{\alpha -1} \mathcal{K}_{\alpha -3}(1)}{(\alpha-3)(\alpha -2) (\alpha -1)} \\
& -\frac{6^{-\alpha } \!\left(81\ 4^{\alpha } \mathcal{K}_{\alpha -3}(3)+128\ 3^{\alpha } \mathcal{K}_{\alpha -3}(4)\right)}{(\alpha -3) (\alpha -2) (\alpha -1)} \\
&+ \frac{4 \left(-2^{6-\alpha }+2^{\alpha }+2^{\alpha } 3^{4-\alpha }-24\right)}{(\alpha -4) (\alpha -3) (\alpha -2) (\alpha -1)} \Bigg]
\end{aligned}
\end{equation}
\\
\hline
\end{tabular}
\end{table*}

\begin{figure*}[b]
	\centering
	\includegraphics[width = 0.95\columnwidth]{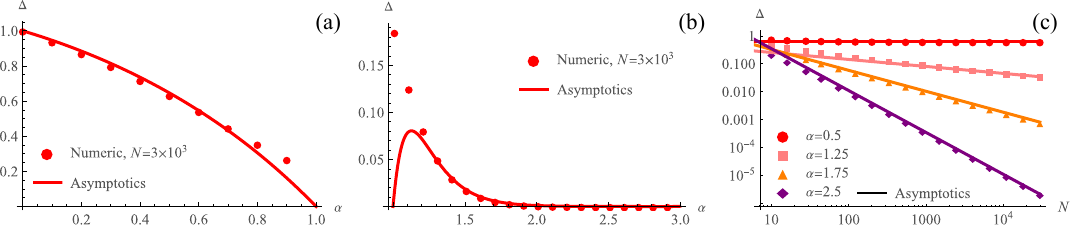}
	\caption{Spectral gap $\Delta_\alpha$ of a one-dimensional hypercube as a function of \textbf{(a)} the power-law exponent $\alpha\in[0,1)$, \textbf{(b)} the power-law exponent $\alpha\in(1,3)$, and \textbf{(c)} the system size $N$, i.e.\ the number of vertices in the hypercube graph. Data points correspond to exact numeric results, while solid curves represent the asymptotic results in Eq.~\eqref{eq:gap_scaling}, with the constants $\mathscr{C}^{(1)}_{i=1,2}$ defined in Table~\ref{tab:spectral_gap_rescaled}, see Eqs.~\eqref{eq:spectral_gap_D1_asymptotic_1} and \eqref{eq:spectral_gap_D1_asymptotic_2}.}
	\label{fig:numeric_vs_asymptotic_d1}
\end{figure*}

\begin{figure*}[h]
	\centering
	\includegraphics[width = 0.95\columnwidth]{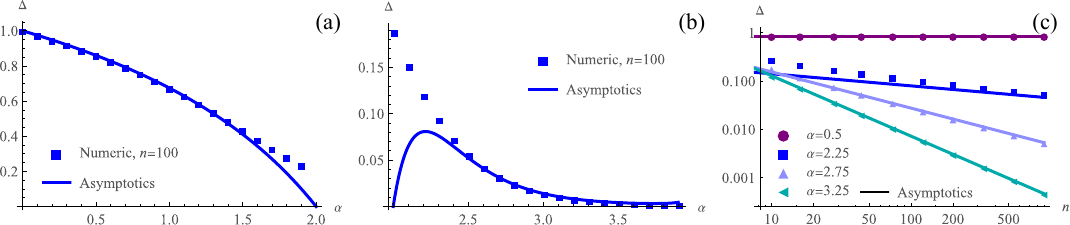}
	\caption{Spectral gap $\Delta_\alpha$ of a two-dimensional hypercubic lattice as a function of \textbf{(a)} the power-law exponent $\alpha\in[0,2)$, \textbf{(b)} the power-law exponent $\alpha\in(2,4)$, and \textbf{(c)} the system size $n=N^{1/2}$, i.e.\ the number of vertices along each dimension in the hypercube graph. Data points correspond to exact numeric results, while solid curves represent the asymptotic results in Eq.~\eqref{eq:gap_scaling} of the main text, with the constants $\mathscr{C}^{(2)}_{i=1,2}$ defined in Table~\ref{tab:spectral_gap_rescaled}, see Eqs.~\eqref{eq:spectral_gap_D2_asymptotic_1} and \eqref{eq:spectral_gap_D2_asymptotic_2}.}
	\label{fig:numeric_vs_asymptotic_d2}
\end{figure*}

\begin{figure*}[h]
	\centering
	\includegraphics[width = 0.95\columnwidth]{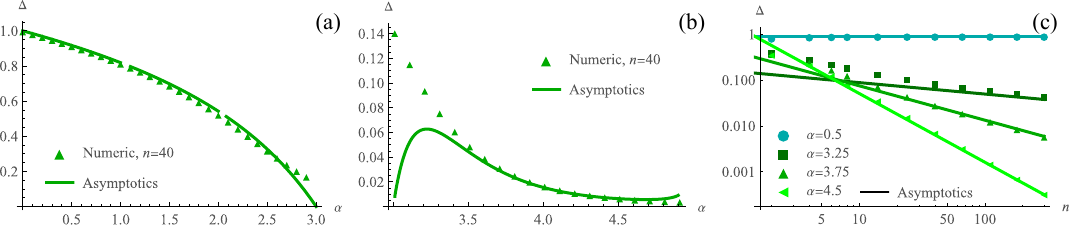}
	\caption{Spectral gap $\Delta_\alpha$ of a three-dimensional hypercube as a function of \textbf{(a)} the power-law exponent $\alpha\in[0,3)$, \textbf{(b)} the power-law exponent $\alpha\in(3,5)$, and \textbf{(c)} the system size $n=N^{1/3}$, i.e.\ the number of vertices along each dimension in the hypercube graph. Data points correspond to exact numeric results, while solid curves represent the asymptotic results in Eq.~\eqref{eq:gap_scaling} of the main text, with the constants $\mathscr{C}^{(3)}_{i=1,2}$ defined in Table~\ref{tab:spectral_gap_rescaled}, see Eqs.~\eqref{eq:spectral_gap_D3_asymptotic_1} and \eqref{eq:spectral_gap_D3_asymptotic_2}.}
	\label{fig:numeric_vs_asymptotic_d3}
\end{figure*}

\begin{figure*}[h]
	\centering
	\includegraphics[width = 0.95\columnwidth]{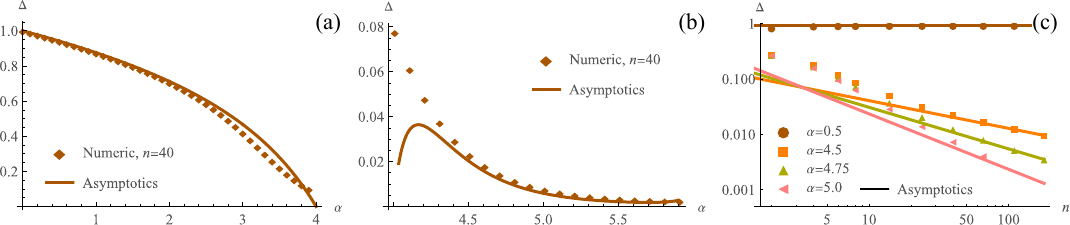}
	\caption{Spectral gap $\Delta_\alpha$ of a four-dimensional hypercube as a function of \textbf{(a)} the power-law exponent $\alpha\in[0,4)$, \textbf{(b)} the power-law exponent $\alpha\in(4,6)$, and \textbf{(c)} the system size $n=N^{1/4}$, i.e.\ the number of vertices along each dimension in the hypercube graph. Data points correspond to exact numeric results, while solid curves represent the asymptotic results in Eq.~\eqref{eq:gap_scaling} of the main text, with the constants $\mathscr{C}^{(3)}_{i=1,2}$ defined in Table~\ref{tab:spectral_gap_rescaled}, see Eqs.~\eqref{eq:spectral_gap_D4_asymptotic_1} and \eqref{eq:spectral_gap_D4_asymptotic_2}.}
	\label{fig:numeric_vs_asymptotic_d4}
\end{figure*}

To benchmark our analytic results for the spectral gap, see Table~\ref{tab:spectral_gap_rescaled}, we perform comparisons with exact numeric results, refer to Figs.~\ref{fig:numeric_vs_asymptotic_d1}--\ref{fig:numeric_vs_asymptotic_d4}. In all cases, subfigure (a) demonstrates the agreement between the analytic spectral gap $\Delta_\alpha\approx 1-\mathscr{C}^{(d)}_1(\alpha)$ with $\alpha\in[0,d)$ and the exact numeric results. Close to the transition point $\alpha\approx d$, finite-size effects become apparent and the accuracy of our analytic approximations deteriorates. To recover the analytic behavior of the gap in the limits $\alpha\rightarrow d^+$ and $\alpha\rightarrow d^-$, system sizes exceeding those that are numerically tractable would be required. In subfigure (b), we provide comparisons for the second asymptotic regime, $\alpha\in(d,d+2)$. Overall, good agreement is observed between analytic and numeric results. As before, toward the edges of this region in $\alpha$ we observe deviations from the analytic predictions. We expect these errors to decrease with increasing $N$, see subfigures (c). Figures \ref{fig:numeric_vs_asymptotic_d1}--\ref{fig:numeric_vs_asymptotic_d4} indicate that the asymptotic results of the main text, together with the constants of Table~\ref{tab:spectral_gap_rescaled}, provide a realistic, quantitative description of $\Delta_\alpha$ in the large-$N$ limit. 

\onecolumngrid


\clearpage
\newpage
\appsection{Detailed derivations of $\delta_\alpha$ and $\mathcal{E}_\alpha(\vec{k}_{\max})$}\label{ss:detailed_derivations}
Here, we provide in-depth calculations of the spectral gap scaling with $N$, see Sec.~\ref{s:spectral_gap_scaling} for the final results. For clarity, we separate the computations according to the hypercubic lattice dimension, with each subsection, \ref{a:ss:d1}--\ref{a:ss:d4}, containing derivations for the constant energy shift $\kappa_0$, the unscaled spectral gap $\delta_\alpha$, and the eigenenergy $\mathcal{E}_\alpha(\vec{k}_{\max})$ with the largest magnitude. 
For our analysis we assume, without loss of generality, that lattices comprise an even number of sites in each dimension, such that $n=N^{1/d}\in \{2x\,\vert\, x\in\mathbb{N}\}$, and we drop the $\alpha$ subscript for convenience. Moreover, the Hurwitz $\zeta$ function $\zeta(\eta, x) = \sum_{n=0}^{\infty} (n + x)^{-\eta}$ for $\text{Re}(\eta) > 1$, and extended by analytic continuation to other $\eta \neq 1$, will be of central importance \cite{Gradshteyn_Ryzhik}. For future reference, we state its large-$x$ asymptotic series expansion here \cite{Nemes17}:
\begin{equation}\label{eq:zeta_expansion}
    \zeta(\eta,x) \sim \frac{x^{-\eta}}{2} + \frac{x^{1-\eta}}{\eta-1}+\mathcal{O}(x^{-1-\eta})\,,
\end{equation}
where $\eta\neq1$ and $x\gg 1$. Finally, note that we use the Manhattan norm in all the derivations of this section. Refer to Sec.~\ref{s:spectral_gap_scaling} for a detailed discussion on this matter. 

\setcounter{subsection}{0}
\subsection{One-dimensional hypercube ($d=1$)}\label{a:ss:d1}
\begin{figure}[b]
    \centering
    \includegraphics[width=\textwidth]{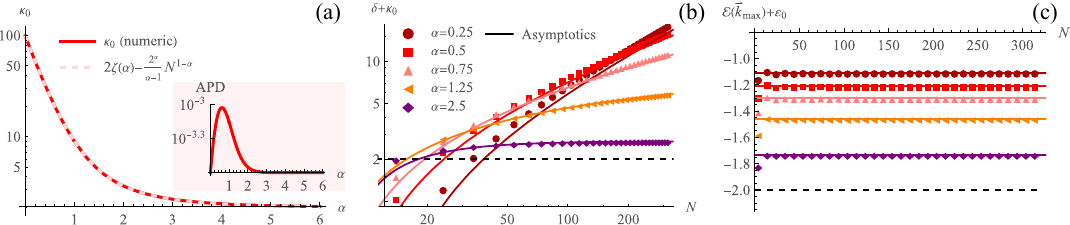}
    \caption{\textbf{(a)} Comparison of the exact numeric value (solid curve) of $\kappa_0$ \eqref{eq:onsite_energy_d1} with the asymptotic result \eqref{eq:a:e0_final_d1} (dashed). Inset shows the absolute percentage deviation, $\rm{APD} = 100\times \vert(E - A)/E \vert$, of the asymptotic result $A$ from the exact result $E$ as a function of $\alpha$. We set $N=100$. \textbf{(b)} Scaling of the spectral gap $\delta$ \eqref{eq:a:delta_Emax_D1} (solid curves) with $N$, and a comparison to exact numeric results (data points). \textbf{(c)} A comparison between analytic \eqref{eq:a:delta_Emax_D1} (solid curves) and exact numeric results (data points) for $\mathcal{E}(\vec{k}_{\max})$. In both (b) and (c) dashed horizontal lines represent the case of nearest-neighbor hopping ($\alpha\rightarrow\infty$).}
    \label{fig:asymptotic_d1}
\end{figure}

\textbf{\textit{Constant energy shift.}} The energy shift $\kappa_0$ is expressed as
\begin{equation}\label{eq:onsite_energy_d1}
	\kappa_0 = \sum_{\substack{j=-N/2+1;\\ j  \neq0}}^{N/2}   \vert j \vert^{-\alpha} = 2 \sum _{j=1}^{N/2} j^{-\alpha }-\left(\frac{N}{2}\right)^{-\alpha } = 2 H_{N/2}^{(\alpha )}-2^{\alpha } N^{-\alpha }\,,
\end{equation}
where $H_N^{(r)}=\sum_{i=1}^N 1/i^r$ is the harmonic number of order $r$ \cite{Gradshteyn_Ryzhik}. For $\alpha\neq1$, the harmonic number in Eq.~\eqref{eq:onsite_energy_d1} can be recast in terms of the Hurwitz and Riemann $\zeta$ functions, leading to $\kappa_0 = 2 \zeta (\alpha) -2 \zeta \left(\alpha,N/2+1\right)-2^{\alpha} N^{-\alpha}$. Asymptotically $\zeta \left(\alpha,N/2+1\right)\approx \zeta \left(\alpha,N/2\right)$, and we use the series expansion of the Hurwitz $\zeta$ function, Eq.~\eqref{eq:zeta_expansion}, to extract the asymptotic behavior of $\kappa_0$ to leading order:
\begin{equation}\label{eq:a:e0_final_d1}
    \kappa_0 \approx 2 \zeta (\alpha) - \frac{2^\alpha}{\alpha-1} N^{1-\alpha} + \mathcal{O}(N^{-\alpha})\,.
\end{equation}
For $\alpha>1$, the $N$-dependent terms decay as a power-law, approaching zero for large $N$, and $\kappa_0$ thus tends toward the real scalar value $2\zeta(\alpha)$. In contrast, for $\alpha<1$, $\kappa_0$ will scale as $\sim N^{1-\alpha}$. This behavior is depicted in Fig.~\ref{fig:asymptotic_d1}(a), together with the absolute percentage deviation (APD) of approximation~\eqref{eq:a:e0_final_d1}. \\

\textbf{\textit{Scaling of $\bm{\delta}$ and $\bm{\mathcal{E}(\vec{k}_{\max})}$.}} A rigorous treatment of $\delta$ and $\mathcal{E}(\vec{k}_{\max})$ is provided in the Supplementary Material of Ref.~\cite{Lewis_etal2021_OptimalQuantumSpatial}. Following their approach, we recover the asymptotic expressions to leading-order as
\begin{equation}\label{eq:a:delta_Emax_D1}
    \delta\approx -\kappa_0 + 2\zeta(\alpha) +2^\alpha  \pi^{\alpha-1} \sin\left(\frac{\pi \alpha}{2}\right)\Gamma(1-\alpha) N^{1-\alpha} + 2\sum_{j=1}^\infty \frac{\zeta(\alpha-2j)}{(2j)!}(2\pi i)^{2j} N^{-2j}\,, \quad \mathcal{E}(\vec{k}_{\max}) \approx -\varepsilon_0 + (2^{2-\alpha}-2)\zeta(\alpha)\,,
\end{equation}
provided $N\gg1$. The accuracy of these expressions is assessed through a comparison to exact numeric results, see Fig.~\ref{fig:asymptotic_d1}.

\subsection{Two-dimensional hypercube ($d=2$)}\label{a:ss:d2}
\begin{figure}[b]
    \centering
    \includegraphics[width=\textwidth]{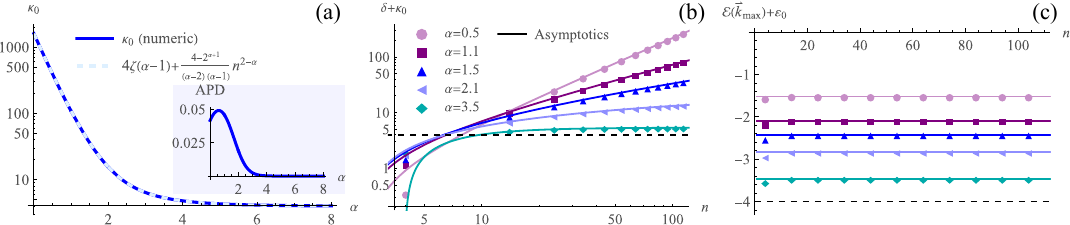}
    \caption{\textbf{(a)} Comparison of the exact numeric value (solid curve) of $\kappa_0$ \eqref{eq:onsite_energy_d2} with the asymptotic result \eqref{eq:a:e0_final_d2} (dashed). Inset shows the absolute percentage deviation, $\rm{APD} = 100\times \vert(E - A)/E \vert$, of the asymptotic result $A$ from the exact result $E$ as a function of $\alpha$. We set $n=N^{1/d}=40$. \textbf{(b)} Scaling of the spectral gap $\delta$ \eqref{eq:spectral_gap_D2_asymptotic_unscaled} (solid curves) with $n$, and a comparison to exact numeric results (data points). \textbf{(c)} A comparison between analytic \eqref{eq:Epi_D2_asymptotic} (solid curves) and exact numeric results (data points) for $\mathcal{E}(\vec{k}_{\max})$. In both (b) and (c) dashed horizontal lines represent the case of nearest-neighbor hopping ($\alpha\rightarrow\infty$).}
    \label{fig:asymptotic_d2}
\end{figure}

\textbf{\textit{Constant energy shift.}} For the two-dimensional hypercubic lattice, $\kappa_0$ is defined as
\begin{equation}\label{eq:onsite_energy_d2}
	\kappa_0 = \sum_{\vec{j} \neq\vec{0}}  1/\vert\vert \vec{j}\vert\vert_1^{\alpha}\ = \hspace{-0.3cm}\sum_{\substack{j_1,j_2=-n/2+1;\\\vert j_1 \vert + \vert j_2 \vert \neq0}}^{n/2}  \hspace{-0.4cm}\left( \vert j_1 \vert + \vert j_2 \vert \right)^{-\alpha},
\end{equation}
where $\vert\vert\dots\vert\vert_1$ denotes the Manhattan norm (geodesic/shortest-path distance), $n=N^{1/d}$ and $\vec{j}\equiv(j_1,j_2)$. Equation \eqref{eq:onsite_energy_d2} can be recast into the following equivalent form:
\begin{equation}\label{eq:onsite_energy_dimension2_reformulated}
	\kappa_0 = \left[1+2^{1+\alpha}(n-1)\right]n^{-\alpha} + \sum_{j=1}^{n/2-1}\left[4j^{1-\alpha}+4j(n-j)^{-\alpha}\right],
\end{equation}
which is amenable to an asymptotic analysis. The leading-order contribution of the first term is $\mathcal{O}(n^{1-\alpha})$. To understand the limiting behavior of the second term of Eq.~\eqref{eq:onsite_energy_dimension2_reformulated} as $n\rightarrow\infty$, we express the summations in terms of the Hurwitz $\zeta$ function $\zeta(\eta,x)$, and then use the asymptotic series expansion of Eq.~\eqref{eq:zeta_expansion}. After some algebra, we obtain
\begin{equation}
	\sum_{j=1}^{n/2-1} 4j^{1-\alpha} = 4\sum_{j=1}^{\infty} j^{1-\alpha} - 4\sum_{j=0}^{\infty} (j+n/2)^{1-\alpha}
	= 4\zeta(\alpha-1) -4 \zeta(\alpha-1,n/2)
	\approx 4\zeta(\alpha-1) - 2^\alpha n^{1-\alpha} - \frac{2^\alpha}{\alpha-2} n^{2-\alpha} + \mathcal{O}(n^{-\alpha})\,,\label{eq:sum_d2_expand_1}
\end{equation}
\begin{align}\label{eq:sum_d2_expand_2}
    \sum_{j=1}^{n/2-1}4j(n-j)^{-\alpha} = 4 \sum_{j=n/2+1}^{n-1}\left[nj^{-\alpha} -j^{1-\alpha}\right]
    &=4 n \left[\zeta \left(\alpha ,\frac{n}{2}+1\right)-\zeta (\alpha ,n)\right] - 4 \left[\zeta \left(\alpha -1,\frac{n}{2}+1\right)-\zeta (\alpha -1,n)\right]\\
    &\approx 2^{\alpha } n^{1-\alpha } + \frac{2^{\alpha } (\alpha -3)+4 }{(\alpha -2) (\alpha -1)}n^{2-\alpha } +\mathcal{O}(n^{-\alpha})\,.\nonumber
\end{align}
Substituting the results of Eqs.~\eqref{eq:sum_d2_expand_1} and \eqref{eq:sum_d2_expand_2} into Eq.~\eqref{eq:onsite_energy_dimension2_reformulated} leads to
\begin{equation}\label{eq:a:e0_final_d2}
    \kappa_0\approx 4\zeta(\alpha-1) + \frac{4-2^{\alpha+1}}{(\alpha-2)(\alpha-1)} n^{2-\alpha} +\mathcal{O}(n^{1-\alpha})\,.
\end{equation}
As observed for the one-dimensional case in Sec.~\ref{a:ss:d1}, there are two important regimes: (i) When $\alpha<2$, the $n$-dependent term dominates such that $\kappa_0\sim  n^{2-\alpha}$, and (ii) for $\alpha>2$, $\kappa_0$ is approximated by the real constant $4\zeta(\alpha-1)$. Due to the suppression of contributions from the terms scaling as $\sim n^x$, $x<0$, this approximation improves as $n$ becomes larger. The accuracy of the asymptotic result \eqref{eq:a:e0_final_d2} is assessed in Fig.~\ref{fig:asymptotic_d2}(a). \\

\textbf{\textit{Scaling of $\bm{\delta}$ and $\bm{\mathcal{E}(\vec{k}_{\max})}$.}} By definition, the unscaled spectral gap is
\begin{equation}
    \delta =-\kappa_0 + \sum_{\vec{j} \neq\vec{0}}  \cos(\vec{k}_1\cdot\vec{j})/\vert\vert \vec{j}\vert\vert_1^{\alpha}\ = -\kappa_0 + \hspace{-0.3cm}\sum_{\substack{j_1,j_2=-n/2+1;\\\vert j_1\vert + \vert j_2\vert\neq 0}}^{n/2} \hspace{-0.3cm} (\vert j_1\vert + \vert j_2\vert)^{-\alpha} \cos\left(2\pi j_1/n\right),
\end{equation}
which can be written in the equivalent form:
\begin{align}
    \delta &= -\kappa_0 + 4 \sum_{j_1,j_2=1}^{n/2}\frac{\cos\left(2\pi j_1/n\right)}{ (j_1+j_2)^{\alpha}} + 2\sum_{j_2=1}^{n/2}j_2^{-\alpha} + 2\sum_{j_2=1}^{n/2}\left(\frac{n}{2}+j_2\right)^{-\alpha}  + \hspace{-0.3cm}\sum_{\substack{j_1=-n/2+1;\\j_1\neq0}}^{n/2} \hspace{-0.3cm}\frac{\cos\left(2\pi j_1/n\right)}{ \vert j_1\vert^{\alpha}} - \hspace{-0.3cm}\sum_{\substack{j_1=-n/2+1;\\j_1\neq0}}^{n/2} \hspace{-0.2cm}\frac{\cos\left(2\pi j_1/n\right) }{ \left(\vert j_1\vert +\frac{n}{2}\right)^{\alpha}}\nonumber\\
    &\approx -\kappa_0 + 4\zeta(\alpha) + \underbrace{4 \sum_{j_1,j_2=1}^{n/2}\cos\left(2\pi j_1/n\right) (j_1+j_2)^{-\alpha}}_{(*)} +\ \mathcal{O}(n^{1-\alpha})\,. \label{eq:alg_connectivity_D2_intermediate_step2}
\end{align}
Simplification was performed by noting that the final two terms, treated in the one-dimensional case, give $2\zeta(\alpha)$ and terms of order $n^{1-\alpha}$. Now, focussing on the term denoted by $(*)$ in Eq.~\eqref{eq:alg_connectivity_D2_intermediate_step2}, we recast the summation exactly as
\begin{equation}\label{eq:alg_connectivity_D2_intermediate_step3}
    (*) = 4 \sum_{j_1,j_2=1}^{n/2}\cos\left(2\pi j_1/n\right) (j_1+j_2)^{-\alpha} = 4 \sum_{j_1=1}^{n/2}\cos\left(2\pi j_1/n\right) \left[\zeta(\alpha,j_1+1)-\zeta(\alpha,j_1+n/2+1) \right].
\end{equation}
To evaluate the finite sum~\eqref{eq:alg_connectivity_D2_intermediate_step3} containing the Hurwitz $\zeta$ functions, we approximate it in terms of an integral and subsequently use the machinery of calculus to extract the asymptotic result. For this purpose, we implement the Euler–Maclaurin formula
\begin{equation}\label{eq:Euler_Maclaurin}
    \sum_{k=a+1}^{b} f(k) = \int_{a}^{b} f(x) \, dx + \frac{1}{2}[f(a) - f(b)] + \sum_{k=1}^{\lfloor p/2\rfloor} \frac{B_{2k}}{(2k)!} \left( f^{(2k-1)}(b) - f^{(2k-1)}(a) \right) + R_p\,,
\end{equation}
where $a,b$ are natural numbers and $f:\mathbb{R}\rightarrow\mathbb{R}$ is a continuous function for real $k\in[a,b]$. Provided $f(k)$ is $p$ times continuously differentiable on the interval $[a,b]$ for $p\in\mathbb{Z}^+$, the difference between the summation and integral depends on terms containing the $k$-th Bernoulli number, $B_k$, as well as the error term $R_p$. Disregarding the error terms, which we later numerically show to have negligible impact on the final results, see Fig.~\ref{fig:asymptotic_d2}(b), the summation of \eqref{eq:alg_connectivity_D2_intermediate_step3} is approximated as
\begin{equation}\label{eq:alg_connectivity_D2_intermediate_step4}
    (*) \approx 4\zeta(\alpha-1)-4\zeta(\alpha) + 4 \int_0^{n/2}dx\, \cos(2\pi x/n) \zeta(\alpha,x+1) - 4 \int_0^{n/2}dx\, \cos(2\pi x/n) \zeta(\alpha,x+n/2+1)\,.
\end{equation}
Performing a change of variable $x\rightarrow ny/2$, followed by a series expansion of the Hurwitz $\zeta$ function~\eqref{eq:zeta_expansion}, Eq.~\eqref{eq:alg_connectivity_D2_intermediate_step4} simplifies to
\begin{equation}\label{eq:alg_connectivity_D2_intermediate_step5}
    (*) \approx 4\zeta(\alpha-1)-4\zeta(\alpha) +  \frac{2^\alpha}{\alpha-1}  \left[\int_0^{1}dy\, \cos(\pi y) \,y^{1-\alpha} -  \int_0^{1}dy\, \cos(\pi y) \,(y+1)^{1-\alpha}\right] n^{2-\alpha} + \mathcal{O}(n^{1-\alpha})\,,
\end{equation}
where only leading-order terms are explicitly written. The integrals \eqref{eq:alg_connectivity_D2_intermediate_step5} can be computed analytically, and written in terms of the generalized hypergeometric function $_pF_q(a;b;z)$ \cite{Gradshteyn_Ryzhik} and the exponential integral function $E_n(x)$ \cite{Gradshteyn_Ryzhik} as
\begin{equation}\label{eq:D2_integral_1}
   \int_0^{1}dy\, \cos(\pi y) \,y^{1-\alpha} = -\frac{\, _1F_2\left(1-\frac{\alpha }{2};\frac{1}{2},2-\frac{\alpha }{2};-\frac{\pi ^2}{4}\right)}{\alpha -2}\,, \quad {\rm for} \quad \alpha<2\,,
\end{equation}
\begin{equation}\label{eq:D2_integral_2}
    \int_0^{1}dy\, \cos(\pi y) \,(y+1)^{1-\alpha}= -\frac{1}{2} \left[E_{\alpha -1}(-i \pi )+E_{\alpha -1}(i \pi )-2^{2-\alpha } \{E_{\alpha -1}(-2 i \pi )+E_{\alpha -1}(2 i \pi )\}\right].
\end{equation}
For the first integral we require $\alpha<2$ to ensure convergence. This restriction on the range of $\alpha$ is not problematic, since the $\mathcal{O}(n^{2-\alpha})$ term of Eq.~\eqref{eq:alg_connectivity_D2_intermediate_step5}, to which the integrals are a prefactor, is asymptotically suppressed when $\alpha>2$ and the term $(*)$ tends to the constant $4\zeta(\alpha-1)-4\zeta(\alpha)$. Thus, in the regime where the $\mathcal{O}(n^{2-\alpha})$ term is dominant, the integral is guaranteed to converge. Finally, Eq.~\eqref{eq:alg_connectivity_D2_intermediate_step5}, together with the intregrals \eqref{eq:D2_integral_1} and \eqref{eq:D2_integral_2}, is substituted back into Eq.~\eqref{eq:alg_connectivity_D2_intermediate_step2} to give the scaling of $\delta$, see Eq.~\eqref{eq:spectral_gap_D2_asymptotic_unscaled}. This asymptotic result is in close agreement with exact numeric results for large $n$, see Fig.~\ref{fig:asymptotic_d2}(b).\\

To determine the scaling of $\mathcal{E}(\vec{k}_{\max})$, we start with the definition 
\begin{equation}
    \mathcal{E}(\vec{k}_{\max}) =-\varepsilon_0 +\sum_{\vec{j} \neq\vec{0}}  \cos(\vec{k}_{\max}\cdot\vec{j})/\vert\vert \vec{j}\vert\vert_1^{\alpha}\ = -\varepsilon_0 + \hspace{-0.3cm}\sum_{\substack{j_1,j_2=-n/2+1;\\\vert j_1\vert + \vert j_2\vert\neq 0}}^{n/2} \hspace{-0.3cm} (\vert j_1\vert + \vert j_2\vert)^{-\alpha} \cos\left(\pi j_1 + \pi j_2\right).
\end{equation}
Since $j_1+j_2\in \mathbb{Z}$, we perform the replacement $\cos\left(\pi j_1 + \pi j_2\right)\rightarrow(-1)^{j_1+j_2}$ and shift the summation indices such that the absolute values may be dropped. After some algebra, we arrive at
\begin{equation}\label{eq:largest_ev_expansion_1}
    \mathcal{E}(\vec{k}_{\max}) = -\varepsilon_0 + \sum_{\substack{j_1,j_2=0;\\j_1+j_2\neq0}}^{n/2}(-1)^{j_1+j_2}(j_1+j_2)^{-\alpha} + 2\sum_{j_1=0}^{n/2}\sum_{j_2=1}^{n/2-1}  (-1)^{j_1-j_2}(j_1+j_2)^{-\alpha} + \sum_{j_1,j_2=1}^{n/2-1}(-1)^{j_1+j_2}(j_1+j_2)^{-\alpha}\,.
\end{equation}
The second term of Eq.~\eqref{eq:largest_ev_expansion_1} can be recast in terms of two summations, each running over a single summation index:
\begin{equation}\label{eq:largest_ev_expansion_B5}
    \sum_{\substack{j_1,j_2=0;\\j_1+j_2\neq0}}^{n/2}(-1)^{j_1+j_2}(j_1+j_2)^{-\alpha} = \sum_{j=1}^{n/2}j^{-\alpha}(-1)^j(j+1) + \sum_{j=1}^{n/2}(n-j+1)^{-\alpha}(-1)^{n-j+1}j \approx (2^{2-\alpha}-1)\zeta(\alpha-1) + (2^{1-\alpha}-1)\zeta(\alpha)\,,
\end{equation}
where the final step assumes $n\gg1$. The third term~\eqref{eq:largest_ev_expansion_1} can be treated similarly. To leading-order, we find
\begin{align}\label{eq:largest_ev_expansion_2}
    2\sum_{j_1=0}^{n/2}\sum_{j_2=1}^{n/2-1}  \frac{(-1)^{j_1-j_2}}{(j_1+j_2)^{\alpha}} &= 2\sum_{j=1}^{n/2-1} j^{-\alpha}(-1)^j j + 2\left(\frac{2}{n}\right)^{\alpha}(-1)^{n/2}\left(\frac{n}{2}-1\right) + 2 \sum_{j=1}^{n/2-1} (n-j)^{-\alpha}(-1)^{n-j} j\nonumber\\
    &= (2^{3-\alpha}-2)\zeta(\alpha-1) +(-1)^{n/2}\left(2^\alpha n^{1-\alpha}-2^{1+\alpha}n^{-\alpha} \right) + 2\sum_{j=n/2+1}^{n-1}j^{-\alpha}(-1)^j(n-j)\\
    &\approx (2^{3-\alpha}-2)\zeta(\alpha-1) + \mathcal{O}(n^{1-\alpha})\,,\label{eq:largest_ev_expansion_3}
\end{align}
where the second term of line \eqref{eq:largest_ev_expansion_2} is sub-leading-order, and the third term is negligible when $n$ is large. The latter statement can be verified by writing the summation explicitly in terms of Hurwitz $\zeta$ functions, $\zeta(\eta,x)$, yielding
\begin{align}\label{eq:B8}
    2\sum_{j=n/2+1}^{n-1}j^{-\alpha}(-1)^j(n-j) =\ & 2^{2-\alpha } i^n \left[ \zeta \left(\alpha -1,\frac{n}{4}+\frac{1}{2}\right)- \zeta \left(\alpha -1,\frac{n}{4}+1\right)+e^{\frac{i \pi  n}{2}} \bigg\{ \zeta \left(\alpha -1,\frac{n}{2}\right)- \zeta \left(\alpha -1,\frac{n}{2}+\frac{1}{2}\right) \right.\nonumber\\
    &\qquad\qquad\left.-\frac{n}{2} \zeta \left(\alpha ,\frac{n}{2}\right)+\frac{n}{2} \zeta \left(\alpha ,\frac{n}{2}+\frac{1}{2}\right)\bigg\}-\frac{n}{2} \zeta \left(\alpha ,\frac{n}{4}+\frac{1}{2}\right)+\frac{n}{2}\zeta \left(\alpha ,\frac{n}{4}+1\right)\right].
\end{align}
Now, note that $n/4+1/2\approx n/4+1$ and $n/2\approx n/2+1/2$ for large $n$. This implies that, asymptotically, $\zeta \left(\eta,\frac{n}{4}+\frac{1}{2}\right)\approx \zeta \left(\eta,\frac{n}{4}+1\right)$ and $\zeta \left(\eta,\frac{n}{2}\right)\approx \zeta \left(\eta,\frac{n}{2}+\frac{1}{2}\right)$ for both $\eta=\alpha$ and $\eta=\alpha-1$, and the entire summation tends to zero in large-$n$ limit. Determining the asymptotic behavior of the final term of Eq.~\eqref{eq:largest_ev_expansion_1} follows the procedure taken in Eq.~\eqref{eq:largest_ev_expansion_B5}. We simply state the result as
\begin{equation}\label{eq:largest_ev_expansion_4}
    \sum_{j_1,j_2=1}^{n/2-1}(-1)^{j_1+j_2}(j_1+j_2)^{-\alpha}\approx (2^{2-\alpha}-1)\zeta(\alpha-1) - (2^{1-\alpha}-1)\zeta(\alpha)\,.
\end{equation}
Substituting Eqs.~\eqref{eq:largest_ev_expansion_B5}, \eqref{eq:largest_ev_expansion_3} and \eqref{eq:largest_ev_expansion_4} into Eq.~\eqref{eq:largest_ev_expansion_1}, we recover the asymptotic result \eqref{eq:Epi_D2_asymptotic}, provided in Table~\ref{tab:spectral_gap_unscaled}. Figure \ref{fig:asymptotic_d2}(c) provides an assessment of the accuracy of this asymptotic result for $\varepsilon_0=0$ and a range of $\alpha$ values.

\subsection{Three-dimensional hypercube ($d=3$)}\label{a:ss:d3}
\begin{figure}[t]
    \centering
    \includegraphics[width=\textwidth]{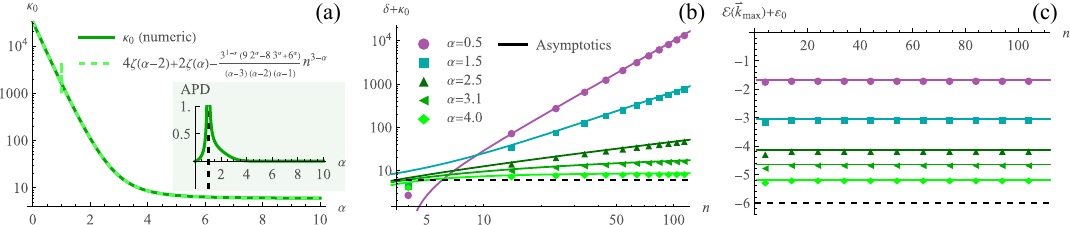}
    \caption{\textbf{(a)} Comparison of the exact numeric value (solid curve) of $\kappa_0$ \eqref{eq:onsite_energy_d3} with the asymptotic result \eqref{eq:a:e0_final_d3} (dashed). 
    Inset shows the absolute percentage deviation, $\rm{APD} = 100\times \vert(E - A)/E \vert$, of the asymptotic result $A$ from the exact result $E$ as a function of $\alpha$. We set $n=N^{1/3}=32$. \textbf{(b)} Scaling of the spectral gap $\delta$ \eqref{eq:spectral_gap_D3_asymptotic_unscaled} (solid curves) with $n$, and a comparison to exact numeric results (data points). \textbf{(c)} A comparison between analytic \eqref{eq:Epi_D3_asymptotic} (solid curves) and exact numeric results (data points) for $\mathcal{E}(\vec{k}_{\max})$. In both (b) and (c) 
    dashed horizontal lines represent the case of nearest-neighbor hopping ($\alpha\rightarrow\infty$).}
    \label{fig:asymptotic_d3}
\end{figure}

\textbf{\textit{Constant energy shift.}} In hypercubic lattices of three dimensions, $\kappa_0$ may be treated via an expansion into a series of summations, followed by various approximations and simplifications. Recalling that $\vert\vert\dots\vert\vert_1$ denotes the Manhattan norm, $n=N^{1/3}$ and $\vec{j}\equiv(j_1,j_2,j_3)$, the definition of the energy $\kappa_0$ leads to
\begin{align}\label{eq:onsite_energy_d3}
	\hspace{-0.1cm}\kappa_0 = \sum_{\vec{j} \neq\vec{0}}  1/\vert\vert \vec{j}\vert\vert_1^{\alpha}\ = \hspace{-0.4cm}\sum_{\substack{j_1,j_2,j_3=-n/2+1;\\\vert j_1 \vert + \vert j_2 \vert + \vert j_3 \vert \neq0}}^{n/2} \hspace{-0.4cm} & \left( \vert j_1 \vert + \vert j_2 \vert + \vert j_3 \vert\right)^{-\alpha} = \ 8\hspace{-0.2cm}\sum_{j_1,j_2,j_3=1}^{n/2-1}(j_1+j_2+j_3)^{-\alpha} + 12\sum_{j_1,j_2=1}^{n/2-1}\left(j_1+j_2+\frac{n}{2}\right)^{-\alpha} + 12\sum_{j_1,j_2=1}^{n/2-1}\left(j_1+j_2\right)^{-\alpha} \nonumber\\
 & \qquad\qquad\qquad\qquad+6\sum_{j_1=1}^{n/2-1}\left(j_1+n\right)^{-\alpha} + 12\sum_{j_1=1}^{n/2-1}\left(j_1+\frac{n}{2}\right)^{-\alpha} + 6\sum_{j_1=1}^{n/2-1}\left(j_1\right)^{-\alpha} +\mathcal{O}(n^{-\alpha})\,.
\end{align}
We proceed by performing an asymptotic expansion of each term in Eq.~\eqref{eq:onsite_energy_d3} independently, after which we combine the results to obtain the scaling of $\kappa_0$ \eqref{eq:kappa0_D3_asymptotic}. Starting with the first term in \eqref{eq:onsite_energy_d3}, we reformulate the triple summation over indices $j_1$, $j_2$ and $j_3$ as a collection of summations with single indices:
\begin{align}\label{eq:onsite_energy_d3_term1}
    8\hspace{-0.2cm}\sum_{j_1,j_2,j_3=1}^{n/2-1}(j_1+j_2+j_3)^{-\alpha} =& \underbrace{4\sum_{j=1}^{n/2+1}(j-1)(j-2)j^{-\alpha}}_{(a)} +\underbrace{\sum_{j=n/2+2}^{n/2+n/4}\left[ -8 j^2+12 j n-3 n (n+2)+8\right] j^{-\alpha}}_{(c)}\\
    & + \underbrace{4\sum_{j=1}^{n/2-1}j(j+1)\left[\frac{3n}{2}-j-2\right]^{-\alpha}}_{(b)}+ \underbrace{\sum_{j=n/2}^{n/2+n/4-3} \left[6 (2 j+3) n-8 (j+1) (j+3)-3 n^2\right] \left[\frac{3 n}{2}-j-2\right]^{-\alpha }}_{(d)}.\nonumber
\end{align}
Note that the summation upper bounds in $(c)$ and $(d)$ contain the fraction $n/4$. For convenience, we hereafter assume that $\rm{\sf mod}$$(n,4)=0$, where $\rm{\sf mod}$ is the standard modulo operation giving the remainder on division of $n$ by $4$. In the limit of large $n$, our results apply to lattices with any number of sites. Now, terms $(a)$ and $(c)$ can be rewritten exactly in terms of the difference of infinite sums, giving expressions in terms of the Hurwitz $\zeta$ function $\zeta(\eta,x)$ and Riemann $\zeta$ function $\zeta(\eta)$. After some algebra, these terms exhibit the following asymptotic behavior:
\begin{align}\label{eq:d3(a)}
    (a) &=\, -4 \left[\zeta \left(\alpha -2,\frac{n}{2}+2\right)-3 \zeta \left(\alpha -1,\frac{n}{2}+2\right)+2 \zeta \left(\alpha ,\frac{n}{2}+2\right)-\zeta (\alpha -2)+3 \zeta (\alpha -1)-2 \zeta (\alpha )\right]\nonumber\\
    &\approx 4 \zeta (\alpha -2)-12 \zeta (\alpha -1)+8 \zeta (\alpha ) -2^{\alpha +2} n^{-\alpha } + \frac{2^{\alpha } (3 \alpha -7) }{\alpha -1} n^{1-\alpha } -\frac{2^{\alpha -1} (\alpha -8) }{\alpha -2}n^{2-\alpha } -\frac{2^{\alpha -1} }{\alpha -3}n^{3-\alpha } + \mathcal{O}(n^{-1-\alpha})\,,
\\
    (c) &=\, -8 \zeta \left(\alpha -2,\frac{n}{2}+2\right)+8 \zeta \left(\alpha -2,\frac{3 n}{4}+1\right)+12 n \zeta \left(\alpha -1,\frac{n}{2}+2\right)-12 n \zeta \left(\alpha -1,\frac{3 n}{4}+1\right)\nonumber\\
    &\quad \quad -(3 n (n+2)-8) \left[\zeta \left(\alpha ,\frac{n}{2}+2\right)-\zeta \left(\alpha ,\frac{3 n}{4}+1\right)\right]\nonumber\\
    &\approx \,2^{\alpha +2} 3^{-\alpha } \left(3^{\alpha }-2^{\alpha }\right) n^{-\alpha }
    +\frac{\left(\frac{2}{3}\right)^{\alpha } \left[3\ 2^{\alpha } (\alpha -3)+3^{\alpha } (7-3 \alpha )\right] }{\alpha -1}n^{1-\alpha }
    -\frac{2^{\alpha -2} 3^{-\alpha } \left(3\ 2^{\alpha }-2\ 3^{\alpha }\right) (\alpha -7) }{\alpha -1} n^{2-\alpha }\nonumber\\
    &\quad \quad 
    +\frac{2^{\alpha -3} 3^{-\alpha } \left[4\ 3^{\alpha } (\alpha -4) (\alpha +1)-9\ 2^{\alpha } (\alpha -5) \alpha \right] }{(\alpha -3) (\alpha -2) (\alpha -1)} n^{3-\alpha } + \mathcal{O}(n^{-1-\alpha})\,.
\end{align}
To evaluate terms $(b)$ and $(d)$ of Eq.~\eqref{eq:onsite_energy_d3_term1} we shift the summation indices, such that the denominator undergoes the transformation $\left[\frac{3 n}{2}-j-2\right]^{-\alpha }\longrightarrow j^{-\alpha}$. In their new form, summations $(b)$ and $(d)$  can be written as the differences of the Hurwitz $\zeta$ function infinite summations. Applying this, we find the asymptotic expansions 
\begin{align}\label{eq:d3(b)}
    (b) &= \sum_{j=n-1}^{3n/2-3}  (2 j-3 n+2) (2 j-3 n+4) j^{-\alpha } = -4 \zeta \left(\alpha -2,\frac{3 n}{2}-2\right)+4 \zeta (\alpha -2,n-1)+12 (n-1) \zeta \left(\alpha -1,\frac{3 n}{2}-2\right)\nonumber\\
    &\qquad \qquad \qquad \qquad \qquad\qquad \qquad\qquad\qquad -12 (n-1) \zeta (\alpha -1,n-1)+(3 n-4) (3 n-2) \left[\zeta (\alpha ,n-1)-\zeta \left(\alpha ,\frac{3 n}{2}-2\right)\right] \nonumber\\
    &\approx  2^2\ 3^{-\alpha } \left(3^{\alpha }-2^{\alpha }\right) n^{-\alpha }
    -\frac{3^{-\alpha } \left[3^{\alpha } (3 \alpha -11)+3\ 2^{\alpha +2}\right]}{\alpha -1}n^{1-\alpha }
    +\frac{3^{-\alpha } \left[3^{\alpha } (\alpha -10) (\alpha -5)-27\ 2^{\alpha +1}\right]}{2 (\alpha -2) (\alpha -1)} n^{2-\alpha }\nonumber\\
    &\quad +\frac{3^{-\alpha } \left(3^{\alpha } ((\alpha -9) \alpha +26)-27\ 2^{\alpha }\right) }{(\alpha -3) (\alpha -2) (\alpha -1)} n^{3-\alpha } + \mathcal{O}(n^{-1-\alpha})\,,
   \end{align}\begin{align}
 \label{eq:d3(d)}
    (d) &= \sum_{j=3n/4+1}^{n-2} \left[-8 j^2+12 j n-3 n (n+2)+8\right] j^{-\alpha } = -8 \zeta \left(\alpha -2,\frac{3 n}{4}+1\right)+8 \zeta (\alpha -2,n-1)+12 n \zeta \left(\alpha -1,\frac{3 n}{4}+1\right)\nonumber\\
    &\qquad \qquad \qquad \qquad \qquad\qquad \qquad\qquad\qquad\qquad-12 n \zeta (\alpha -1,n-1)-(3 n (n+2)-8) \left[\zeta \left(\alpha ,\frac{3 n}{4}+1\right)-\zeta (\alpha ,n-1)\right] \nonumber\\
    &\approx 2^2\ 3^{-\alpha } \left(4^{\alpha }-3^{\alpha }\right) n^{-\alpha } 
    + \frac{3^{-\alpha } \left[3^{\alpha } (3 \alpha -11)-3\ 4^{\alpha } (\alpha -3)\right] }{\alpha -1} n^{1-\alpha }
    + \frac{3^{-\alpha } \left[3\ 4^{\alpha } (\alpha -7)-2\ 3^{\alpha } (\alpha -13)\right] }{4 (\alpha -1)} n^{2-\alpha } \nonumber\\
    &\quad
    + \frac{3^{-\alpha } \left[9\ 4^{\alpha } (\alpha -5) \alpha -8\ 3^{\alpha } (\alpha^2 -9\alpha +2)\right] }{8 (\alpha -3) (\alpha -2) (\alpha -1)} n^{3-\alpha }+ \mathcal{O}(n^{-1-\alpha})\,.
\end{align} 
Combing the results of Eqs.~\eqref{eq:d3(a)}--\eqref{eq:d3(d)}, we arrive at an asymptotic expression for the first term of Eq.~\eqref{eq:onsite_energy_d3}:
\begin{equation}\label{eq:d3_first_term_final_asymptotic_result}
   8\sum_{j_1,j_2,j_3=1}^{n/2-1}(j_1+j_2+j_3)^{-\alpha} \approx 4 \zeta (\alpha -2)-12 \zeta (\alpha -1)+8 \zeta (\alpha ) -\frac{3^{1-\alpha } \left(9\ 2^{\alpha }-8\ 3^{\alpha }+6^{\alpha }\right) }{(\alpha -3) (\alpha -2) (\alpha -1)}n^{3-\alpha } + \mathcal{O}(n^{2-\alpha})\,,
\end{equation}
where we keep only leading-order terms in $n$ and constants. Determining the large-$n$ behavior of the remaining terms in Eq.~\eqref{eq:onsite_energy_d3} is more straightforward. For the second term we find
\begin{align}
    12\sum_{j_1,j_2=1}^{n/2-1}\left(j_1+j_2+\frac{n}{2}\right)^{-\alpha} &= 12\sum_{j=1}^{n/2} (j-1) \left(j+\frac{n}{2}\right)^{-\alpha }  + 12\sum_{j=1}^{n/2-2}j \left(\frac{3 n}{2}-j-1\right)^{-\alpha }\nonumber\\
    &= 12\left[\zeta \left(\alpha -1,\frac{3 n}{2}-1\right)-\zeta (\alpha -1,n+1)+\left(1-\frac{3 n}{2}\right)\left[ \zeta \left(\alpha ,\frac{3 n}{2}-1\right)- \zeta (\alpha ,n+1)\right]\right.\nonumber\\
    & \quad \left.+2^{\alpha } n^{-\alpha }+\zeta \left(\alpha -1,\frac{n}{2}\right)-\zeta (\alpha -1,n+1)-\frac{1}{2} (n+2) \left[\zeta \left(\alpha ,\frac{n}{2}\right)-\zeta (\alpha ,n+1)\right]\right]\nonumber\\
    & \approx -\frac{2^{\alpha +1} 3^{1-\alpha } \left(3^{\alpha }-3\right) }{\alpha -1}n^{1-\alpha }
    + \frac{3^{1-\alpha } \left(9\ 2^{\alpha }-8\ 3^{\alpha }+6^{\alpha }\right) }{\left(\alpha-2\right)(\alpha-1)}n^{2-\alpha } + \mathcal{O}(n^{-\alpha})\,.\label{eq:d3_second_term_final_asymptotic_result}
\end{align}
Since the leading-order contribution scales as $\sim n^{3-\alpha}$, see Eq.~\eqref{eq:d3_first_term_final_asymptotic_result}, the result above indicates that the contribution of the second term to $\kappa_0$, see Eq.~\eqref{eq:onsite_energy_d3}, is negligible for $n\gg1$. We can determine the asymptotic behavior of the third term in Eq.~\eqref{eq:onsite_energy_d3} following a similar sequence of steps:
\begin{align}
    12\sum_{j_1,j_2=1}^{n/2-1}\left(j_1+j_2\right)^{-\alpha} &= 12\sum_{j=1}^{n/2} (j-1) j^{-\alpha }  + 12\sum_{j=n/2+1}^{n-2}(n-j-1) j^{-\alpha } \nonumber\\
    &= -12 \left[2 \zeta \left(\alpha -1,\frac{n}{2}+1\right)-\zeta (\alpha -1,n-1)-n \zeta \left(\alpha ,\frac{n}{2}+1\right)+(n-1) \zeta (\alpha ,n-1)-\zeta (\alpha -1)+\zeta (\alpha )\right]\nonumber\\
    &\approx 12 \zeta (\alpha -1)-12 \zeta (\alpha ) + \frac{12 }{\alpha -1} n^{1-\alpha } + \frac{6 \left(2^{\alpha }-2\right) }{(\alpha-2)(\alpha-1)}n^{2-\alpha } + \mathcal{O}(n^{-\alpha})\,.\label{eq:d3_third_term_final_asymptotic_result}
\end{align}
Crucially, the expansion contains two $n$-independent terms, which will contribute to the value of $\kappa_0$ when $\alpha>3$. The dominant $n$-dependent term in the expansion scales as $\sim n^{2-\alpha}$, inferring that the contribution will be sub-leading-order in the final expansion. The final three terms in~\eqref{eq:onsite_energy_d3} can each be directly reformulated in terms of the difference of two appropriate infinite sums. This allows us to express the summations in terms of the well-known Hurwitz $\zeta$ function, and expand using the asymptotic result of Eq.~\eqref{eq:zeta_expansion}, as before. The calculations are summarized below:
\begin{align}
    &\sum_{j_1=1}^{n/2-1}\frac{6}{\left(j_1+n\right)^{\alpha}} = \sum_{j_1=0}^{\infty}\frac{6}{\left(j_1+n+1\right)^{\alpha}} - 6\sum_{j_1=0}^{\infty}\left(j_1+\frac{3n}{2}\right)^{-\alpha} = 6 \zeta (\alpha ,n+1)-6 \zeta \left(\alpha ,\frac{3 n}{2}\right) \approx  \frac{3^{1-\alpha } \left(2\ 3^{\alpha }-3\ 2^{\alpha }\right) n^{1-\alpha }}{\alpha -1} + \mathcal{O}( n^{-\alpha})\,,\label{eq:d3_fourth_term_final_asymptotic_result}\\
    &12\sum_{j_1=1}^{n/2-1}\left(j_1+\frac{n}{2}\right)^{-\alpha} = 12\left[\sum_{j_1=0}^{\infty}\left(j_1+\frac{n}{2}+1\right)^{-\alpha} - \sum_{j_1=0}^{\infty}\left(j_1+n\right)^{-\alpha} \right] \label{eq:d3_fifth_term_final_asymptotic_result}= 12 \left[\zeta \left(\alpha ,\frac{n}{2}+1\right)- \zeta (\alpha ,n) \right] \approx \frac{6 \left(2^{\alpha }-2\right) }{\alpha -1} n^{1-\alpha } + \mathcal{O}( n^{-\alpha})\,,\\
    &6\sum_{j_1=1}^{n/2-1}\left(j_1\right)^{-\alpha} = 6\sum_{j_1=1}^{\infty}\left(j_1\right)^{-\alpha} - 6\sum_{j_1=0}^{\infty}\left(j_1+\frac{n}{2}\right)^{-\alpha} = 6 \zeta (\alpha )-6 \zeta \left(\alpha ,\frac{n}{2}\right) \approx 6 \zeta (\alpha )- \frac{6\ 2^{\alpha -1} }{\alpha -1} n^{1-\alpha } +\mathcal{O}( n^{-\alpha})\,.\label{eq:d3_sixth_term_final_asymptotic_result}
\end{align}
The scaling $n^{1-\alpha}$ is sub-leading-order in all of the above expressions. We therefore disregard these subdominant contributions in the full expansion of $\kappa_0$ where the leading-order contribution scales as $\sim n^{3-\alpha}$. Finally, combining Eqs.~\eqref{eq:d3_first_term_final_asymptotic_result}--\eqref{eq:d3_sixth_term_final_asymptotic_result}, the energy $\kappa_0$ \eqref{eq:onsite_energy_d3} takes the form
\begin{equation}\label{eq:a:e0_final_d3}
   \kappa_0 \approx 4 \zeta (\alpha -2)+2 \zeta (\alpha ) -\frac{3^{1-\alpha } \left(9\ 2^{\alpha }-8\ 3^{\alpha }+6^{\alpha }\right) }{(\alpha -3) (\alpha -2) (\alpha -1)}n^{3-\alpha } + \mathcal{O}(n^{2-\alpha})\,,
\end{equation}
as reported in Table~\ref{tab:spectral_gap_unscaled}. 
For $\alpha<3$, $\kappa_0\sim n^{3-\alpha}$, meanwhile $\kappa_0$ is well-approximated by the $\alpha$-dependent constant $4\zeta(\alpha-2)+2\zeta(\alpha)$ in the regime $\alpha>3$. Figure~\ref{fig:asymptotic_d3}(a) provides an indication of how well the asymptotics of Eq.~\eqref{eq:a:e0_final_d3} describe $\kappa_0$, even for comparatively small $n$. \\

\textbf{\textit{Scaling of $\bm{\delta}$ and $\bm{\mathcal{E}(\vec{k}_{\max})}$.}} For three spatial dimensions, the (unnormalized) spectral gap is given by
\begin{equation}\label{eq:alg_connectivity_D3_intermediate_step1}
    \delta = -\kappa_0 + \sum_{\vec{j} \neq\vec{0}}  \cos(\vec{k}_1\cdot\vec{j})/\vert\vert \vec{j}\vert\vert_1^{\alpha}\ = -\kappa_0 + \hspace{-0.4cm}\sum_{\substack{j_1,j_2,j_3=-n/2+1;\\\vert j_1\vert + \vert j_2\vert + \vert j_3\vert\neq 0}}^{n/2} \hspace{-0.4cm} (\vert j_1\vert + \vert j_2\vert +\vert j_3\vert)^{-\alpha} \cos\left(2\pi j_1/n\right).
\end{equation}
Omitting intermediate steps, we shift the summation indices in Eq.~\eqref{eq:alg_connectivity_D3_intermediate_step1} such that $j_i\in\mathbb{Z}>0$ for $i=1,2,3$, leading to
\begin{equation}\label{eq:alg_connectivity_D3_intermediate_step2}
    \delta = -\kappa_0 + 12\zeta(\alpha-1)-6\zeta(\alpha)+ \underbrace{8\hspace{-0.2cm}\sum_{j_1,j_2,j_3=1}^{n/2-1}\hspace{-0.2cm}\cos(2\pi j_1/n) (j_1+j_2+j_3)^{-\alpha}}_{(*)} +\ \mathcal{O}(n^{2-\alpha})\,.
\end{equation}
A complete understanding of the scaling requires an asymptotic analysis of the term denoted by $(*)$. Since the argument of the cosine is independent of $j_{i=2,3}$, we can compute
\begin{align}\label{eq:alg_connectivity_D3_intermediate_step3}
    \sum_{j_2,j_3=1}^{n/2-1} (j_1+j_2+j_3)^{-\alpha} &= \sum_{j=1}^{n/2}(j-1)(j+j_1)^{-\alpha} + \sum_{j=1}^{n/2-2}j(n-1-j+j_1)^{-\alpha}  \nonumber\\
    &=\zeta (\alpha -1,j_1+2)-(j_1+1) \zeta (\alpha ,j_1+2)-2 \zeta \left(\alpha -1,j_1+\frac{n}{2}+1\right)+\zeta (\alpha -1,j_1+n-1)\nonumber\\
    &\qquad +(2 j_1+n) \zeta \left(\alpha ,j_1+\frac{n}{2}+1\right)-(j_1+n-1) \zeta (\alpha ,j_1+n-1)\,.
\end{align}
Evidently, Eqs.~\eqref{eq:alg_connectivity_D3_intermediate_step2} and \eqref{eq:alg_connectivity_D3_intermediate_step3} lead to
\begin{align}\label{eq:D3_AG_star}
    (*) = 8\sum_{j_1=1}^{n/2-1} \cos\left(\frac{2\pi j_1}{n}\right) &\bigg[ \zeta (\alpha -1,j_1+2)-(j_1+1) \zeta (\alpha ,j_1+2)-2 \zeta \left(\alpha -1,j_1+\frac{n}{2}+1\right)+\zeta (\alpha -1,j_1+n-1)\nonumber\\
    &\quad +(2 j_1+n) \zeta \left(\alpha ,j_1+\frac{n}{2}+1\right)-(j_1+n-1) \zeta (\alpha ,j_1+n-1) \bigg]\,.
\end{align}
The scaling analysis is identical for each term of the above expression, therefore we simply state the steps, followed by the results for the six terms. As in the two-dimensional case, see Sec.~\ref{a:ss:d2}, we first approximate the summation by an integral using the Euler-Maclaurin formula~\eqref{eq:Euler_Maclaurin}. Disregarding the error terms, we perform a change of variable $j_1\rightarrow ny/2$, followed by the Hurwitz $\zeta$ series expansion \eqref{eq:zeta_expansion}. With some algebra, we obtain the six asymptotic results, written to leading-order in $n$:
\begin{subequations}\label{eq:six_asymp_results}
\begin{align}\label{eq:AG_D3_asymp1}
    8\sum_{j_1=1}^{n/2-1} \cos\left(\frac{2\pi j_1}{n}\right)\zeta (\alpha -1,j_1+2) &\sim 8\zeta(\alpha-2)-8\zeta(\alpha-1) + \frac{2^\alpha}{\alpha-2}\left[\int_0^1 dy\,\cos(\pi y)\ y^{2-\alpha} \right] n^{3-\alpha}\,,\\
    -8\sum_{j_1=1}^{n/2-1} \cos\left(\frac{2\pi j_1}{n}\right) (j_1+1) \zeta (\alpha ,j_1+2)  &\sim -4\zeta(\alpha-2) -4\zeta(\alpha-1)+8\zeta(\alpha) - \frac{2^\alpha}{\alpha-1}\left[\int_0^1 dy\,\cos(\pi y)\ y^{2-\alpha} \right] n^{3-\alpha}\,,\\
   -16 \sum_{j_1=1}^{n/2-1} \cos\left(\frac{2\pi j_1}{n}\right)\zeta \left(\alpha -1,j_1+\frac{n}{2}+1\right) &\sim  -\frac{2^{\alpha+1}}{\alpha-2} \left[ \int_0^1 dy\,\cos(\pi y)\ (y+1)^{2-\alpha}\right] n^{3-\alpha}\,,\\
  8 \sum_{j_1=1}^{n/2-1} \cos\left(\frac{2\pi j_1}{n}\right)\zeta (\alpha -1,j_1+n-1) &\sim  \frac{2^{\alpha}}{\alpha-2} \left[ \int_0^1 dy\,\cos(\pi y)\ (y+2)^{2-\alpha}\right] n^{3-\alpha}\,,\\
    8\sum_{j_1=1}^{n/2-1} \cos\left(\frac{2\pi j_1}{n}\right) (2 j_1+n) \zeta \left(\alpha ,j_1+\frac{n}{2}+1\right) &\sim \frac{2^{\alpha+1}}{\alpha-1} \left[ \int_0^1 dy\,\cos(\pi y)\ (y+1)^{2-\alpha}\right] n^{3-\alpha}\,,\\\label{eq:AG_D3_asymp6}
   - 8\sum_{j_1=1}^{n/2-1} \cos\left(\frac{2\pi j_1}{n}\right) (j_1+n-1) \zeta (\alpha ,j_1+n-1) &\sim -\frac{2^\alpha}{\alpha-1} \left[ \int_0^1 dy\,\cos(\pi y)\ (y+2)^{2-\alpha}  \right] n^{3-\alpha}\,.
\end{align}
\end{subequations}
The integrals \eqref{eq:six_asymp_results}, contributing to the $\alpha$-dependent prefactors of the $\mathcal{O}(n^{3-\alpha})$ terms, are computed exactly, with
\begin{subequations}
\begin{align}
    \int_0^1 dy\,\cos(\pi y)\ y^{2-\alpha} &= -\frac{\, _1F_2\left(\frac{3}{2}-\frac{\alpha }{2};\frac{1}{2},\frac{5}{2}-\frac{\alpha }{2};-\frac{\pi ^2}{4}\right)}{\alpha-3 }\,, \quad \mathrm{for} \quad \alpha<3\,,
\\
    \int_0^1 dy\,\cos(\pi y)\ (y+1)^{2-\alpha} &= -2^{-1}\mathcal{K}_{\alpha -2}(1) + 2^{2-\alpha } \mathcal{K}_{\alpha-2}(2)\,,
\\
    \int_0^1 dy\,\cos(\pi y)\ (y+2)^{2-\alpha}&=2^{2-\alpha } \mathcal{K}_{\alpha -2}(2)-2^{-1} 3^{3-\alpha } \mathcal{K}_{\alpha -2}(3)\,.
\end{align}\label{eq:int_prefactor_d3}
\end{subequations}
Here we use the compact notation $\mathcal{K}_n(z)\equiv E_n(i\pi z) + E_n(-i\pi z)$, with $E_n(\pm i\pi z)$ the exponential integral function \cite{Gradshteyn_Ryzhik}. Substituting the asymptotic results \eqref{eq:six_asymp_results}, together with the prefactors \eqref{eq:int_prefactor_d3}, into Eq.~\eqref{eq:D3_AG_star} leads to
\begin{equation}
    (*) \approx 4 \zeta (\alpha -2)-12 \zeta (\alpha -1)+8 \zeta (\alpha ) - \Bigg[\frac{2^{\alpha }  \, _1F_2\left(\frac{3-\alpha}{2};\frac{1}{2},\frac{5-\alpha}{2};-\frac{\pi ^2}{4}\right)}{(\alpha -3) (\alpha -2) (\alpha -1)} - \frac{2^{\alpha } \mathcal{K}_{\alpha -2}( 1)-4 \mathcal{K}_{\alpha -2}( 2)-2^{\alpha -1} 3^{3-\alpha } \mathcal{K}_{\alpha -2}(3)}{(\alpha -2) (\alpha -1)}  \Bigg] n^{3-\alpha }\,.
\end{equation}
Finally, by replacing the term denoted by $(*)$ in Eq.~\eqref{eq:alg_connectivity_D3_intermediate_step2} by the above expression, we determine the scaling of the spectral gap, see Eq.~\eqref{eq:spectral_gap_D3_asymptotic_unscaled}. Refer to Fig.~\ref{fig:asymptotic_d3}(b) for a comparison of this asymptotic result~\eqref{eq:spectral_gap_D3_asymptotic_unscaled} to exact numeric results.\\

The final quantity to evaluate is $\mathcal{E}(\vec{k}_{\max})$. We start from the definition
\begin{equation}\label{eq:largest_ev_D3_step1}
    \mathcal{E}(\vec{k}_{\max}) =-\varepsilon_0 +\sum_{\vec{j} \neq\vec{0}}  \cos(\vec{k}_{\max}\cdot\vec{j})/\vert\vert \vec{j}\vert\vert_1^{\alpha}\ = -\varepsilon_0 + \hspace{-0.4cm}\sum_{\substack{j_1,j_2,j_3=-n/2+1;\\\vert j_1\vert + \vert j_2\vert+ \vert j_3\vert\neq 0}}^{n/2} \hspace{-0.4cm} (\vert j_1\vert + \vert j_2\vert+ \vert j_3\vert)^{-\alpha} \cos\left(\pi j_1 + \pi j_2 + \pi j_3\right).
\end{equation}
Noting that $\cos\left(\pi j_1 + \pi j_2+\pi j_3\right)=(-1)^{j_1+j_2+j_3}$ for $j_1+j_2+j_3\in \mathbb{Z}$, we expand Eq.~\eqref{eq:largest_ev_D3_step1}:
\begin{align}\label{eq:largest_ev_D3_step2}
    \mathcal{E}(\vec{k}_{\max}) =& -\varepsilon_0 + \hspace{-0.4cm}\sum_{\substack{j_1,j_2,j_3=0;\\j_1 + j_2+ j_3\neq 0}}^{n/2} \hspace{-0.4cm} (j_1+j_2+j_3)^{-\alpha} (-1)^{j_1+j_2+j_3} +  \hspace{-0.3cm}\sum_{j_1,j_2,j_3=1}^{n/2-1} \hspace{-0.3cm} (j_1+j_2+j_3)^{-\alpha} (-1)^{j_1+j_2+j_3}\\
    &+ \underbrace{3\sum_{j_1,j_2=0}^{n/2}\sum_{j_3=1}^{n/2-1}(j_1+j_2+j_3)^{-\alpha} (-1)^{j_1+j_2-j_3} + 3\sum_{j_1=0}^{n/2}\sum_{j_2,j_3=1}^{n/2-1}(j_1+j_2+j_3)^{-\alpha} (-1)^{j_1-j_2-j_3}}_{\mathcal{A}}\,. \nonumber
\end{align}
To analytically extract the scaling, we express the first and second term of Eq.~\eqref{eq:largest_ev_D3_step2} as a series of summations that contain only a single summation index. An identical strategy is implemented in the treatment of two-dimensional hypercubic lattices, see Sec.~\ref{a:ss:d2}. We obtain
\begin{align}\label{eq:largest_ev_D3_step3}
     \hspace{-0.4cm}\sum_{\substack{j_1,j_2,j_3=0;\\j_1 + j_2+ j_3\neq 0}}^{n/2} \hspace{-0.4cm} (j_1+j_2+j_3)^{-\alpha} (-1)^{j_1+j_2+j_3} =& \sum_{x=1}^{n/2} \frac{1}{2}(-1)^x (x^2+3x+2) x^{-\alpha} + \sum_{x=1}^{n/2+1}\frac{1}{2}(-1)^{3n/2-x+1}x (x+1) \left(\frac{3n}{2}-x+1\right)^{-\alpha} \nonumber\\
     &+ \sum_{x=1}^{n/2-1}(-1)^{n/2+x}\left[-x^2+\frac{n}{2}x +\frac{1}{8}(n+2)(n+4)\right]\left(\frac{n}{2}+x\right)^{-\alpha}\,,
\end{align}
and similarly for the second term of Eq.~\eqref{eq:largest_ev_D3_step2}. From the more transparent structure of Eq.~\eqref{eq:largest_ev_D3_step3}, it is possible to derive the asymptotic results
\begin{align}\label{eq:largest_ev_D3_step4}
   \sum_{\substack{j_1,j_2,j_3=0;\\j_1 + j_2+ j_3\neq 0}}^{n/2} \hspace{-0.4cm} (j_1+j_2+j_3)^{-\alpha} (-1)^{j_1+j_2+j_3} &\approx\left(2^{2-\alpha}-\frac{1}{2}\right)\zeta(\alpha-2) + \left(3\ 2^{1-\alpha}-\frac{3}{2}\right)\zeta(\alpha-1) + (2^{1-\alpha}-1)\zeta(\alpha)\,,\\
   \label{eq:largest_ev_D3_step5}
    \sum_{j_1,j_2,j_3=1}^{n/2-1} \hspace{-0.3cm} (j_1+j_2+j_3)^{-\alpha} (-1)^{j_1+j_2+j_3}&\approx \left(2^{2-\alpha}-\frac{1}{2}\right)\zeta(\alpha-2) - \left(3\ 2^{1-\alpha}-\frac{3}{2}\right)\zeta(\alpha-1) + (2^{1-\alpha}-1)\zeta(\alpha)\,.
\end{align}
In fact, this asymptotic behavior comes solely from the first term of \eqref{eq:largest_ev_D3_step3}, since the other terms tend toward zero for an increasing number of lattice sites $n$. See, for example, the calculation of Eq.~\eqref{eq:B8} for the two-dimensional lattice and the discussion thereof. Now, we reformulate the final two terms of Eq.~\eqref{eq:largest_ev_D3_step2}, denoted by $\mathcal{A}$, as
\begin{equation}
    \mathcal{A} = \frac{3}{2}\sum_{x=1}^{n/2-1} x^{-\alpha}(-1)^x x(x+1)+ \frac{3}{2}\sum_{x=1}^{n/2} x^{-\alpha}(-1)^x x(x-1) + C\,,
\end{equation}
where $C\rightarrow 0$ for large $n$. As a result, $\mathcal{A}$ can be approximated by
\begin{equation}\label{eq:calligrpahic_A}
    \mathcal{A}\approx -3\ 2^{-\alpha } \left[\left(2^{\alpha }-8\right) \zeta (\alpha -2)+i^n \left(a+b\right)\right],
\end{equation}
with $a=2 \zeta \left(\alpha -2,\frac{n}{4}\right)-4 \zeta \left(\alpha -2,\frac{n}{4}+\frac{1}{2}\right)+2 \zeta \left(\alpha -2,\frac{n}{4}+1\right)$ and $b=\zeta \left(\alpha -1,\frac{n}{4}\right)-\zeta \left(\alpha -1,\frac{n}{4}+1\right)$. As argued in Sec.~\ref{a:ss:d2}, for very large $n$, we have that $n/4\approx n/4+1/2\approx n/4+1$. Asymptotically, the second term of Eq.~\eqref{eq:calligrpahic_A} then goes to zero, such that $\mathcal{A}\approx (3\ 2^{3-\alpha} - 3) \zeta(\alpha-2)$. Substituting the asymptotic results \eqref{eq:largest_ev_D3_step4}, \eqref{eq:largest_ev_D3_step5} and $\mathcal{A}$ into the expression~\eqref{eq:largest_ev_D3_step2} for $\mathcal{E}(\vec{k}_{\max})$, we get
\begin{equation}\label{eq:largestEVD3_final_app}
    \mathcal{E}(\vec{k}_{\max}) \approx -\varepsilon_0 + (2^{5-\alpha}-4)\zeta(\alpha-2)+(2^{2-\alpha}-2)\zeta(\alpha)\,.
\end{equation}
In Fig.~\ref{fig:asymptotic_d3}(c) we demonstrate that this asymptotic result~\eqref{eq:largestEVD3_final_app} captures the true behavior of the largest eigenenergy well.

\subsection{Four-dimensional hypercube ($d=4$)}\label{a:ss:d4}
\begin{figure}[t]
    \centering
    \includegraphics[width=\textwidth]{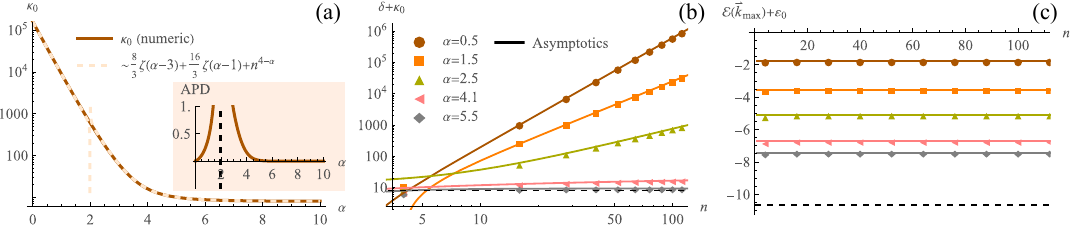}
    \caption{\textbf{(a)} Comparison of the exact numeric value (solid curve) of $\kappa_0$ \eqref{eq:onsite_energy_d4} with the asymptotic result \eqref{eq:a:e0_final_d4} (dashed). Inset shows the absolute percentage deviation, $\rm{APD} = 100\times \vert(E - A)/E \vert$, of the asymptotic result $A$ from the exact result $E$ as a function of $\alpha$. We set $n=N^{1/4}=20$. \textbf{(b)} Scaling of the spectral gap $\delta$ \eqref{eq:spectral_gap_D4_asymptotic_unscaled} (solid curves) with $n$, and a comparison to exact numeric results (data points). \textbf{(c)} A comparison between analytic \eqref{eq:Epi_D4_asymptotic} (solid curves) and exact numeric results (data points) for $\mathcal{E}(\vec{k}_{\max})$. In both (b) and (c) 
    dashed horizontal lines represent the case of nearest-neighbor hopping ($\alpha\rightarrow\infty$).}
    \label{fig:asymptotic_d4}
\end{figure}

\textbf{\textit{Constant energy shift.}} From the definition of $\kappa_0$ we write
\begin{align}\label{eq:onsite_energy_d4}
	\kappa_0 = \sum_{\vec{j} \neq\vec{0}}  1/\vert\vert \vec{j}\vert\vert_1^{\alpha}\ = \hspace{-0.9cm}\sum_{\substack{j_1,j_2,j_3,j_4=-n/2+1;\\\vert j_1 \vert + \vert j_2 \vert + \vert j_3 \vert +\vert j_4 \vert \neq0}}^{n/2} \hspace{-0.9cm}  \left( \vert j_1 \vert + \vert j_2 \vert + \vert j_3 \vert +\vert j_4 \vert\right)^{-\alpha}=& \, 16 \sum _{j_1,j_2,j_3,j_4=1}^{n/2-1}  (j_1+j_2+j_3+j_4)^{-\alpha }+32 \sum _{j_1,j_2,j_3=1}^{n/2-1} \left(j_1+j_2+j_3+\frac{n}{2}\right)^{-\alpha }\nonumber\\
 & +32 \sum _{j_1,j_2,j_3=1}^{n/2-1} (j_1+j_2+j_3)^{-\alpha }+48 \sum _{j_1,j_2=1}^{n/2-1} \left(j_1+j_2+\frac{n}{2}\right)^{-\alpha }\nonumber\\
 &+24 \sum _{j_1,j_2=1}^{n/2-1} (j_1+j_2)^{-\alpha }+24 \sum _{j_1,j_2=1}^{n/2-1}  (j_1+j_2+n)^{-\alpha } + 8 \sum _{j_1=1}^{n/2-1} \left(j_1+\frac{3n}{2}\right)^{-\alpha} \nonumber\\
 &+24\sum _{j_1=1}^{n/2-1} \left(j_1+n\right)^{-\alpha} +24\sum _{j_1=1}^{n/2-1} \left(j_1+\frac{n}{2}\right)^{-\alpha} +8\sum _{j_1=1}^{n/2-1} \left(j_1\right)^{-\alpha} +\mathcal{O}(n^{-\alpha})\,,
\end{align}
where $\vert\vert\dots\vert\vert_1$ denotes the Manhattan norm, $n=N^{1/4}$ is the number of sites in each dimension and $\vec{j}\equiv(j_1,j_2,j_3,j_4)$. The expansion~\eqref{eq:onsite_energy_d4} can be treated on a term-by-term basis. Hereafter, we assume $n\geq14$ for convenience. This does not impact on the validity of the asymptotic result, since we are only interested in the scaling behavior at large $n$. With this assumption, we expand the first term of Eq.~\eqref{eq:onsite_energy_d4}: 
\begin{align}\label{eq:onsite_energy_d4_term1_a_To_d}
     &\hspace{-0.4cm}\sum _{\substack{j_1,j_2,j_3,\\j_4=1}}^{n/2-1} \hspace{-0.1cm} \frac{16}{(j_1+j_2+j_3+j_4)^{\alpha } }\overset{\rm{for}\,\, n \geq 14}{=} 16\Bigg[\underbrace{\sum_{j=1}^{n/2+2} \frac{(j-3) (j-2) (j-1)}{6j^\alpha}  }_{(a)} +\underbrace{\sum_{j=1}^{n/2-1} \frac{j (j+1) (j+2)}{6(-j+2 n-3)^{\alpha }}  }_{(b)}\\
    &\quad\quad+ \underbrace{\frac{1}{2}\sum_{j=1}^{n/2-2} \frac{\left(-j^3+j^2 \left(\frac{n}{2}-4\right)+j \left(\frac{n^2}{4}-3\right)+\frac{n}{24} \left(n^2-4\right)\right)}{ \left(j+\frac{n}{2}+2\right)^{\alpha }}}_{(c)} +\underbrace{\frac{1}{2}\sum_{j=1}^{n/2-3} \frac{\left(-j^3+j^2 \left(\frac{n}{2}-4\right)+j \left(\frac{n^2}{4}-3\right)+\frac{n}{24} \left(n^2-4\right)\right)}{ \left(-j+\frac{3 n}{2}-2\right)^{\alpha }}}_{(d)}\Bigg]\nonumber.
\end{align}
After some manipulation, terms $(a)$--$(d)$ can be rewritten exactly in terms of the differences of infinite sums. These infinite sums have the structure of either the Hurwitz $\zeta$ function $\zeta(\eta,x)$ or the Riemann $\zeta$ function $\zeta(\eta)$, offering further simplification. The asymptotic behavior is then extracted by applying the series expansion of Eq.~\eqref{eq:zeta_expansion}. After some algebra, term $(a)$ of Eq.~\eqref{eq:onsite_energy_d4_term1_a_To_d} exhibits the following asymptotic behavior:
\begin{align}\label{eq:aympD4_termA}
    (a) &=  \bigg[\zeta \left(\alpha -2,\frac{n}{2}+3\right)-\frac{1}{6}\zeta \left(\alpha -3,\frac{n}{2}+3\right)-\frac{11}{6} \zeta \left(\alpha -1,\frac{n}{2}+3\right)+ \zeta \left(\alpha ,\frac{n}{2}+3\right) +\frac{1}{6}\zeta (\alpha -3)- \zeta (\alpha -2)+\frac{11}{6} \zeta (\alpha -1)- \zeta (\alpha )\bigg]\nonumber\\
    &\approx \frac{\zeta (\alpha -3)}{6}-\zeta (\alpha -2)+\frac{11 \zeta (\alpha -1)}{6}-\zeta (\alpha ) 
    + \frac{2^{\alpha -3} (23-11 \alpha ) }{3 (\alpha -1)}n^{1-\alpha }
    + \frac{2^{\alpha -3} (3 \alpha -17) }{3 (\alpha -2)}n^{2-\alpha }\nonumber\\
    &\quad   -\frac{2^{\alpha -5} (\alpha -15) }{3 (\alpha -3)}n^{3-\alpha }
    -\frac{2^{\alpha -5} }{3(\alpha-4) }n^{4-\alpha }
    + \mathcal{O}(n^{-\alpha})\,.
\end{align}
Similarly, we determine the asymptotics of the remaining three terms as
\begin{align}\label{eq:aympD4_termB}
    (b)&= \sum _{j=3n/2-2}^{2 n-4} \frac{(-j+2 n-3) (-j+2 n-2) (-j+2 n-1)}{6} j^{-\alpha } \nonumber\\
    &=\frac{1}{6} \bigg[-\zeta \left(\alpha -3,\frac{3 n}{2}-2\right)+\zeta (\alpha -3,2 n-3) +6 (n-1) \left[\zeta \left(\alpha -2,\frac{3 n}{2}-2\right)- \zeta (\alpha -2,2 n-3)\right]\nonumber\\
    &\quad -(12 (n-2) n+11) \left[\zeta \left(\alpha -1,\frac{3 n}{2}-2\right)- \zeta (\alpha -1,2 n-3)\right] +2 (n-1) (2 n-3) (2 n-1) \left[\zeta \left(\alpha ,\frac{3 n}{2}-2\right)- \zeta (\alpha ,2 n-3)\right]\bigg]\nonumber\\
    &\approx \frac{2^{-\alpha -3} 3^{-\alpha -1} \left[4^{\alpha } (11 \alpha -47)+16\ 3^{\alpha +1}\right] }{\alpha -1} n^{1-\alpha }
    +   \frac{2^{-\alpha -3} 3^{-\alpha -1} \left[176\ 3^{\alpha }-3\ 4^{\alpha } (\{\alpha -14\} \alpha +57)\right]}{(\alpha -2) (\alpha -1)}n^{2-\alpha }\nonumber\\
   &\quad  +\frac{\left[4^{\alpha } (\alpha  \{(\alpha -42) \alpha +407\}-1518)+512\ 3^{\alpha +1}\right]}{2^{\alpha +5} 3^{\alpha +1} (\alpha -3) (\alpha -2) (\alpha -1)}n^{3-\alpha }+\frac{ \left[4^{\alpha } (\alpha  \{(\alpha -18) \alpha +143\}-510)+512\ 3^{\alpha }\right] }{2^{\alpha +5} 3^{\alpha }(\alpha -4) (\alpha -3) (\alpha -2) (\alpha -1)}n^{4-\alpha }
    + \mathcal{O}(n^{-\alpha})\,,
\end{align}
\begin{align}\label{eq:aympD4_termC}
    (c) &= \sum _{j=n/2+3}^n \frac{1}{2} j^{-\alpha } \left[\left(\frac{n^2}{4}-3\right) \left(j-\frac{n}{2}-2\right)-\left(j-\frac{n}{2}-2\right)^3+\left(\frac{n}{2}-4\right) \left(j-\frac{n}{2}-2\right)^2+\frac{n}{24}  \left(n^2-4\right)\right]\nonumber\\
    &= \frac{1}{2} \left[\zeta (\alpha -3,n+1)-\zeta \left(\alpha -3,\frac{n}{2}+3\right)\right] + (n+1) \left[\zeta \left(\alpha -2,\frac{n}{2}+3\right)-\zeta (\alpha -2,n+1)\right] \nonumber\\
    &\quad -\frac{1}{2} (n (n+4)-1) \left[\zeta \left(\alpha -1,\frac{n}{2}+3\right)-\zeta (\alpha -1,n+1)\right] +\frac{1}{12} (n (n+2) (n+4)-12) \left[\zeta \left(\alpha ,\frac{n}{2}+3\right)-\zeta (\alpha ,n+1)\right]\nonumber\\
    &\approx \frac{\left[\left(11\ 2^{\alpha }-14\right) \alpha -23\ 2^{\alpha }+38\right] }{24 (\alpha -1)}n^{1-\alpha }+ \frac{\left[2^{\alpha } (\alpha  ((\alpha -18) \alpha +47)+18)-4 (\alpha -5) ((\alpha -13) \alpha +6)\right]}{96 (\alpha -3) (\alpha -2) (\alpha -1)} n^{3-\alpha }
    \nonumber\\
    &\quad+ \frac{\left[6 \alpha ^2-46 \alpha -2^{\alpha } (\alpha -5) (3 \alpha -5)+56\right] }{24 (\alpha -2) (\alpha -1)}n^{2-\alpha }
    +\frac{\left[-8 (\alpha -7) (\alpha -2) \alpha +2^{\alpha } (\alpha -5) ((\alpha -1) \alpha +6)+96\right] }{96 (\alpha -4) (\alpha -3) (\alpha -2) (\alpha -1)}n^{4-\alpha }
    + \mathcal{O}(n^{-\alpha})\,,
\end{align}
\begin{align}\label{eq:aympD4_termD}
    (d) &= \sum _{j=n+1}^{3n/2-3} \frac{1}{2} j^{-\alpha } \left[\left(\frac{n^2}{4}-3\right) \left(-j+\frac{3 n}{2}-2\right)-\left(-j+\frac{3 n}{2}-2\right)^3+\left(\frac{n}{2}-4\right) \left(-j+\frac{3 n}{2}-2\right)^2+\frac{n}{24}  \left(n^2-4\right)\right]\nonumber\\
    &= \frac{1}{2} \left[\zeta (\alpha -3,n+1)-\zeta \left(\alpha -3,\frac{3 n}{2}-2\right)\right] -(2 n-1) \left[\zeta (\alpha -2,n+1)-\zeta \left(\alpha -2,\frac{3 n}{2}-2\right)\right] \nonumber\\
    &\qquad +\frac{(n-1) (5 n+1)}{2}  \left[\zeta (\alpha -1,n+1)-\zeta \left(\alpha -1,\frac{3 n}{2}-2\right)\right] -\frac{(n (n (11 n-6)-20)+12)}{12}  \left[\zeta (\alpha ,n+1)-\zeta \left(\alpha ,\frac{3 n}{2}-2\right)\right]\nonumber\\
    &\approx \frac{3^{-\alpha -1} \left[2\ 3^{\alpha } (7 \alpha -19)+2^{\alpha } (47-11 \alpha )\right] }{8 (\alpha -1)} n^{1-\alpha }
    + \frac{3^{-\alpha -1} \left[3\ 2^{\alpha } (\alpha -11) (\alpha -3)-2\ 3^{\alpha } (\alpha -5) (3 \alpha -8)\right] }{8 (\alpha -2) (\alpha -1)}n^{2-\alpha }\nonumber\\
    &\qquad +\frac{3^{-\alpha -1} \left[4\ 3^{\alpha } (\alpha -5) ((\alpha -13) \alpha +6)+2^{\alpha } (222-\alpha  ((\alpha -42) \alpha +407))\right]}{32 (\alpha -3) (\alpha -2) (\alpha -1)}n^{3-\alpha }\nonumber\\
    &\qquad
    +\frac{3^{-\alpha -1} \left[8\ 3^{\alpha } ((\alpha -7) (\alpha -2) \alpha +60)-3\ 2^{\alpha } (\alpha  ((\alpha -18) \alpha +143)+138)\right]}{32 (\alpha -4) (\alpha -3) (\alpha -2) (\alpha -1)}n^{4-\alpha }
    + \mathcal{O}(n^{-\alpha})\,.
\end{align}
Combining the results of Eqs.~\eqref{eq:aympD4_termA}--\eqref{eq:aympD4_termD} up to $\mathcal{O}(n^{3-\alpha})$, which is the first sub-leading-order contribution to $\kappa_0$ when $d=4$, we find
\begin{align}\label{eq:onsite_energy_term1_final}
    16\hspace{-0.4cm}\sum _{j_1,j_2,j_3,j_4=1}^{n/2-1} \hspace{-0.4cm} (j_1+j_2+j_3+j_4)^{-\alpha } \approx& \frac{8 \zeta (\alpha -3)+88 \zeta (\alpha -1)}{3}-16( \zeta (\alpha -2)+ \zeta (\alpha )) -\frac{ \left(3^{\alpha } \left(-3\ 2^{\alpha +3}+4^{\alpha }-64\right)+81\ 4^{\alpha }\right) }{2^{\alpha-2} 3^{\alpha }(\alpha -4) (\alpha -3) (\alpha -2) (\alpha -1)}n^{4-\alpha }.
\end{align}
Moreover, it can be checked that the second term appearing in Eq.~\eqref{eq:onsite_energy_d4} is $\mathcal{O}(n^{3-\alpha})$ and has no constants, i.e.\ no $n$-independent terms. Therefore, its contribution can immediately be neglected in our asymptotic analysis. For the third term, see Eq.~\eqref{eq:onsite_energy_d4}, we already derived the asymptotic expression in Sec.~\ref{a:ss:d3}. We therefore multiply Eq.~\eqref{eq:d3_first_term_final_asymptotic_result} by $4$ to obtain
\begin{equation}
\label{eq:d4_third_term_final_asymptotic_result}
   32\sum_{j_1,j_2,j_3=1}^{n/2-1}(j_1+j_2+j_3)^{-\alpha} \approx 16 \zeta (\alpha -2)-48 \zeta (\alpha -1)+32 \zeta (\alpha ) -\frac{2^2\ 3^{1-\alpha } \left(9\ 2^{\alpha }-8\ 3^{\alpha }+6^{\alpha }\right) }{(\alpha -3) (\alpha -2) (\alpha -1)}n^{3-\alpha } + \mathcal{O}(n^{2-\alpha})\,.
\end{equation}
Likewise, the large-$n$ asymptotic behavior of the fourth and fifth terms of Eq.~\eqref{eq:onsite_energy_d4} is determined by Eqs.~\eqref{eq:d3_second_term_final_asymptotic_result} and \eqref{eq:d3_third_term_final_asymptotic_result}, up to a constant prefactor. We simply state the results below, refering the reader back to Eqs.~\eqref{eq:d3_second_term_final_asymptotic_result} and \eqref{eq:d3_third_term_final_asymptotic_result} for further details: 
\begin{align}
    48\sum_{j_1,j_2=1}^{n/2-1}\left(j_1+j_2+\frac{n}{2}\right)^{-\alpha}  &\approx -\frac{2^{\alpha +3} 3^{1-\alpha } \left(3^{\alpha }-3\right) }{\alpha -1}n^{1-\alpha }
    + \frac{2^2\ 3^{1-\alpha } \left(9\ 2^{\alpha }-8\ 3^{\alpha }+6^{\alpha }\right) }{\left(\alpha-2\right)(\alpha-1)}n^{2-\alpha } + \mathcal{O}(n^{-\alpha})\,,\label{eq:d4_fourth_term_final_asymptotic_result}\\
    24\sum_{j_1,j_2=1}^{n/2-1}\left(j_1+j_2\right)^{-\alpha} &\approx 24 \zeta (\alpha -1)-24 \zeta (\alpha ) + \frac{24}{\alpha -1} n^{1-\alpha } + \frac{12 \left(2^{\alpha }-2\right) }{(\alpha-2)(\alpha-1)}n^{2-\alpha } + \mathcal{O}(n^{-\alpha})\,.\label{eq:d4_fifth_term_final_asymptotic_result}
\end{align}
In Eq.~\eqref{eq:onsite_energy_d4} the sixth term can be treated similarly, such that
\begin{align}
    &24\sum_{j_1,j_2=1}^{n/2-1}\left(j_1+j_2+n\right)^{-\alpha} = 24\sum_{j=1}^{n/2} (j-1) \left(j+n\right)^{-\alpha }  + 24\sum_{j=1}^{n/2-2}j \left(2n-j-1\right)^{-\alpha }\nonumber\\
    &= 24\bigg[3 n\, \zeta \left(\alpha ,\frac{3 n}{2}+1\right)-2 \zeta \left(\alpha -1,\frac{3 n}{2}+1\right)+\zeta (\alpha -1,n)+\zeta (\alpha -1,2 n-1) -(n+1)\, \zeta (\alpha ,n)+(1-2 n)\, \zeta (\alpha ,2 n-1)+n^{-\alpha }\bigg]\nonumber\\
    & \approx -\frac{3\ 2^{3-\alpha } \left(2^{\alpha }-2\right)}{\alpha -1} n^{1-\alpha }
    + \frac{2^{2-\alpha } 3^{1-\alpha } \left[2\ 3^{\alpha } \left(2^{\alpha }+4\right)-9\ 4^{\alpha }\right]}{(\alpha -2) (\alpha -1)} n^{2-\alpha } + \mathcal{O}(n^{-\alpha})\,.\label{eq:d4_sixth_term_final_asymptotic_result}
\end{align}
Note that the leading-order term of Eq.~\eqref{eq:d4_sixth_term_final_asymptotic_result} is sub-leading-order in the context of the full energy $\kappa_0$, thus making a negligible contribution to $\kappa_0$ in the large-$n$ limit. Finally, we combine the last four terms of $\kappa_0$ \eqref{eq:onsite_energy_d4}, for which the scaling follows directly from the steps provided in Eqs.~\eqref{eq:d3_fourth_term_final_asymptotic_result}--\eqref{eq:d3_sixth_term_final_asymptotic_result} of Sec.~\ref{a:ss:d3}, leading to
\begin{align}\label{eq:d4_onsite_energy_term_final_final}
    &\text{Last 4 terms: }\ 8 \sum _{j_1=1}^{n/2-1} \left(j_1+\frac{3n}{2}\right)^{-\alpha} +24\sum _{j_1=1}^{n/2-1} \left(j_1+n\right)^{-\alpha} +24\sum _{j_1=1}^{n/2-1} \left(j_1+\frac{n}{2}\right)^{-\alpha} +8\sum _{j_1=1}^{n/2-1} \left(j_1\right)^{-\alpha}\nonumber\\
    &= 8 \left[H_{n/2-1}^{(\alpha )}-2 \zeta \left(\alpha ,\frac{3 n}{2}\right)+3 \zeta \left(\alpha ,\frac{n}{2}+1\right)-\zeta (\alpha ,2 n)- \left(\frac{2^\alpha}{3^\alpha}+3\right) n^{-\alpha }\right]\approx 8 \left[\zeta (\alpha )+ \frac{ \left(-2\ 3^{\alpha }-3\ 4^{\alpha }+12^{\alpha }\right) }{6^{\alpha }(\alpha -1)} n^{1-\alpha }\right] + \mathcal{O}(n^{-\alpha})\,,
\end{align}
where only the $\alpha$-dependent constant, $8\zeta (\alpha )$, will enter in the final asymptotic expansion of $\kappa_0$. Using Eqs.~\eqref{eq:onsite_energy_term1_final}, \eqref{eq:d4_third_term_final_asymptotic_result}, \eqref{eq:d4_fifth_term_final_asymptotic_result}, and \eqref{eq:d4_onsite_energy_term_final_final}, we obtain, for large $n$,
\begin{equation}\label{eq:a:e0_final_d4}
    \kappa_0 \approx \frac{8}{3} \zeta (\alpha -3) +\frac{16}{3}\zeta (\alpha -1) -\frac{4 \left(-2^{6-\alpha }+2^{\alpha }+2^{\alpha } 3^{4-\alpha }-24\right)}{(\alpha -4) (\alpha -3) (\alpha -2) (\alpha -1)}n^{4-\alpha } + \mathcal{O}(n^{3-\alpha})\,.
\end{equation}
As observed in Secs.~\ref{a:ss:d1}--\ref{a:ss:d3}, $\alpha=d$ separates two asymptotic regimes, each with distinct behavior. In the $\alpha<4$ regime, $\kappa_0\sim n^{4-\alpha}$. However, when $\alpha>4$, $\kappa_0$ is approximated by an $\alpha$-dependent constant $\frac{8}{3} \zeta (\alpha -3) +\frac{16}{3}\zeta (\alpha -1)$, which naturally becomes more accurate with increasing $n$ due to the suppression of sub-leading-order terms scaling as $\sim n^{x-\alpha}$, $x<4$. For an illustration of the accuracy of the asymptotic result \eqref{eq:a:e0_final_d4}, refer to Fig.~\ref{fig:asymptotic_d4}(a).\\

\textbf{\textit{Scaling of $\bm{\delta}$ and $\bm{\mathcal{E}(\vec{k}_{\max})}$.}} The (unnormalized) spectral gap is expressed as
\begin{equation}\label{eq:alg_connectivity_D4_intermediate_step1}
    \delta = -\kappa_0 + \sum_{\vec{j} \neq\vec{0}}  \cos(\vec{k}_1\cdot\vec{j})/\vert\vert \vec{j}\vert\vert_1^{\alpha}\ = -\kappa_0 + \hspace{-0.7cm}\sum_{\substack{j_1,j_2,j_3,j_4=-n/2+1;\\\vert j_1\vert + \vert j_2\vert + \vert j_3\vert+\vert j_4\vert\neq 0}}^{n/2} \hspace{-0.7cm} (\vert j_1\vert + \vert j_2\vert +\vert j_3\vert+\vert j_4\vert)^{-\alpha} \cos\left(2\pi j_1/n\right).
\end{equation}
Following the same approach as in Sections \ref{a:ss:d2} and \ref{a:ss:d3}, we recast the gap~\eqref{eq:alg_connectivity_D4_intermediate_step1} in a form that only accounts for constants and leading-order contributions in $n$:
\begin{equation}\label{eq:alg_connectivity_D4_intermediate_step2}
    \delta = -\kappa_0 +16\zeta(\alpha-2) -24\zeta(\alpha-1)+16\zeta(\alpha)+ \underbrace{16\hspace{-0.3cm}\sum_{j_1,j_2,j_3,j_4=1}^{n/2-1}\hspace{-0.3cm}\cos(2\pi j_1/n) (j_1+j_2+j_3+j_4)^{-\alpha}}_{(*)} + \mathcal{O}(n^{3-\alpha})\,.
\end{equation}
The quadruple summation, denoted by $(*)$~\eqref{eq:alg_connectivity_D4_intermediate_step2}, is simplified by evaluating the summations with indices $j_{i=2,3,4}$, whereby
\begin{align}\label{eq:alg_connectivity_D4_intermediate_step3}
    8\sum_{\substack{j_2,j_3,j_4=1}}^{n/2-1}(j_1+j_2+j_3+j_4)^{-\alpha}= \,\,&4 \sum _{j=1}^{n/2+1} (j-2) (j-1) (j+j_1)^{-\alpha }+ \sum _{j=n/2+2}^{3n/4} \left(-8 j^2+12 j n-3 n (n+2)+8\right) (j+j_1)^{-\alpha }\nonumber\\
    &+\sum _{j=n/2}^{3n/4-3} \frac{\left[6n (2 j+3) -8 (j+1) (j+3)-3 n^2\right]}{\left(-j+j_1+\frac{3 n}{2}-2\right)^{\alpha}}+4 \sum _{j=1}^{n/2-1} \frac{j (j+1)}{\left(-j+j_1+\frac{3 n}{2}-2\right)^{\alpha}}\,.
\end{align}
Exploiting this representation~\eqref{eq:alg_connectivity_D4_intermediate_step3} in terms of single-index summations, we express each summation as the difference of infinite sums. After some manipulation and shifting of indices, the infinite summations evaluate to Hurwitz zeta functions $\zeta(\eta, x)$ and Eq.~\eqref{eq:alg_connectivity_D4_intermediate_step3} reduces to 
\begin{align}\label{eq:alg_connectivity_D4_intermediate_step4}
   8\hspace{-0.2cm}\sum_{j_2,j_3,j_4=1}^{n/2-1} \hspace{-0.2cm}&(j_1+j_2+j_3+j_4)^{-\alpha}= 4 \zeta (\alpha -2,j_1+1)-4 (2 j_1+3) \zeta (\alpha -1,j_1+1)+4 (j_1+1) (j_1+2) \zeta (\alpha ,j_1+1)\nonumber\\
   &\qquad \quad   -12 \zeta \left(\alpha -2,j_1+\frac{n}{2}+2\right)-4 \zeta \left(\alpha -2,j_1+\frac{3 n}{2}-2\right)+12 \zeta (\alpha -2,j_1+n-1)\nonumber\\
   &\qquad \quad +12 (2 j_1+n+1) \zeta \left(\alpha -1,j_1+\frac{n}{2}+2\right)+4 (2 j_1+3 n-3) \zeta \left(\alpha -1,j_1+\frac{3 n}{2}-2\right)\nonumber\\
   &\qquad \quad -12 (2 j_1+2 n-1) \zeta (\alpha -1,j_1+n-1)-3 (2 j_1+n) (2 j_1+n+2) \zeta \left(\alpha ,j_1+\frac{n}{2}+2\right)\nonumber\\
   &\qquad \quad +(2 j_1+3 n-2) (-2 j_1-3 n+4) \zeta \left(\alpha ,j_1+\frac{3 n}{2}-2\right)+12 (j_1+n-1) (j_1+n) \zeta (\alpha ,j_1+n-1)\,.
\end{align}
Inserting this result into Eq.~\eqref{eq:alg_connectivity_D4_intermediate_step2}, we  now extract the asymptotic behavior of the term labeled by $(*)$. This involves computing the twelve terms independently, and later combining the results. For transparency, we state the leading-order behavior with $n$ for each term below. The approach is identical to that implemented in the preceding sections: Approximate the summations by integrals and perform a change of variable, followed by a large-$n$ expansion of the Hurwitz $\zeta$ function \eqref{eq:zeta_expansion}. Refer to Eqs.~\eqref{eq:AG_D4_asymp1}--\eqref{eq:AG_D4_asymp12} for the final results.
\begin{subequations}
\begin{align}\label{eq:AG_D4_asymp1}
    8&\sum_{j_1=1}^{n/2-1} \cos\left(\frac{2\pi j_1}{n}\right)\zeta (\alpha -2,j_1+1) \approx 8\zeta(\alpha-3)-8\zeta(\alpha-2) + \frac{2^{\alpha-1}}{(\alpha-3)}\left[\int_0^1 dy\,\cos(\pi y)\ y^{3-\alpha} \right] n^{4-\alpha}\,,
    \\
    -8&\sum_{j_1=1}^{n/2-1} \cos\left(\frac{2\pi j_1}{n}\right)(2 j_1+3) \zeta (\alpha -1,j_1+1) \approx 8\left(3\zeta(\alpha-1)-\zeta(\alpha-3)-2\zeta(\alpha-2)\right) - \frac{2^\alpha}{\alpha-2}\left[\int_0^1 dy\,\cos(\pi y)\ y^{3-\alpha} \right] n^{4-\alpha}\,,
\\
    8&\sum_{j_1=1}^{n/2-1} \cos\left(\frac{2\pi j_1}{n}\right) \frac{(j_1+1) (j_1+2)}{[\zeta (\alpha ,j_1+1)]^{-1}} \approx 8\left(\frac{ \zeta (\alpha -3)}{3}+ \zeta (\alpha -2)+\frac{2\zeta (\alpha -1)}{3} -2 \zeta (\alpha ) \right)+ \frac{2^{\alpha-1}}{\alpha-1}\left[\int_0^1 dy\,\cos(\pi y)\ y^{3-\alpha} \right] n^{4-\alpha}\,,
\\
    -24&\sum_{j_1=1}^{n/2-1} \cos\left(\frac{2\pi j_1}{n}\right)\zeta \left(\alpha -2,j_1+\frac{n}{2}+2\right) \approx -\frac{3\ 2^{\alpha-1}}{\alpha-3}\left[\int_0^1 dy\,\cos(\pi y)\ (y+1)^{3-\alpha} \right] n^{4-\alpha}\,,
\\
   -8 &\sum_{j_1=1}^{n/2-1} \cos\left(\frac{2\pi j_1}{n}\right)\zeta \left(\alpha -2,j_1+\frac{3 n}{2}-2\right) \approx -\frac{2^{\alpha-1}}{\alpha-3}\left[\int_0^1 dy\,\cos(\pi y)\ (y+3)^{3-\alpha} \right] n^{4-\alpha}\,,
\\
    24&\sum_{j_1=1}^{n/2-1} \cos\left(\frac{2\pi j_1}{n}\right) \zeta (\alpha -2,j_1+n-1) \approx \frac{3\ 2^{\alpha-1}}{\alpha-3} \left[\int_0^1 dy\,\cos(\pi y)\ (y+2)^{3-\alpha} \right] n^{4-\alpha}\,,
\\
    24&\sum_{j_1=1}^{n/2-1} \cos\left(\frac{2\pi j_1}{n}\right)(2 j_1+n+1) \zeta \left(\alpha -1,j_1+\frac{n}{2}+2\right) \approx \frac{3\ 2^{\alpha}}{\alpha-2} \left[\int_0^1 dy\,\cos(\pi y)\ (y+1)^{3-\alpha} \right] n^{4-\alpha}\,,
\\
    8&\sum_{j_1=1}^{n/2-1} \cos\left(\frac{2\pi j_1}{n}\right)(2 j_1+3 n-3) \zeta \left(\alpha -1,j_1+\frac{3 n}{2}-2\right) \approx \frac{ 2^{\alpha}}{\alpha-2} \left[\int_0^1 dy\,\cos(\pi y)\ (y+3)^{3-\alpha} \right] n^{4-\alpha}\,,
\\
    -24&\sum_{j_1=1}^{n/2-1} \cos\left(\frac{2\pi j_1}{n}\right) (2 j_1+2 n-1) \zeta (\alpha -1,j_1+n-1) \approx -\frac{3\ 2^{\alpha}}{\alpha-2} \left[\int_0^1 dy\,\cos(\pi y)\ (y+2)^{3-\alpha} \right] n^{4-\alpha}\, ,
\\
    -6&\sum_{j_1=1}^{n/2-1} \cos\left(\frac{2\pi j_1}{n}\right) (2 j_1+n) (2 j_1+n+2) \zeta \left(\alpha ,j_1+\frac{n}{2}+2\right)\approx -\frac{3\ 2^{\alpha-1}}{\alpha-1} \left[\int_0^1 dy\,\cos(\pi y)\ (y+1)^{3-\alpha} \right] n^{4-\alpha}\,,
\\
    2&\sum_{j_1=1}^{n/2-1} \cos\left(\frac{2\pi j_1}{n}\right) (2 j_1+3 n-2) (-2 j_1-3 n+4) \zeta \left(\alpha ,j_1+\frac{3 n}{2}-2\right) \approx -\frac{2^{\alpha-1}}{\alpha-1} \left[\int_0^1 dy\,\cos(\pi y)\ (y+3)^{3-\alpha} \right] n^{4-\alpha}\,,
\\ \label{eq:AG_D4_asymp12}
    24&\sum_{j_1=1}^{n/2-1} \cos\left(\frac{2\pi j_1}{n}\right) (j_1+n-1) (j_1+n) \zeta (\alpha ,j_1+n-1) \approx \frac{3\ 2^{\alpha-1}}{\alpha-1} \left[\int_0^1 dy\,\cos(\pi y)\ (y+2)^{3-\alpha} \right] n^{4-\alpha}\,.
\end{align}
\end{subequations}
In the asymptotic results, see Eqs.~\eqref{eq:AG_D4_asymp1}--\eqref{eq:AG_D4_asymp12}, there are several integrals, all contributing to the $\alpha$-dependent prefactors of the terms scaling as $\sim n^{4-\alpha}$. These have analytic solutions, and are given by
\begin{subequations}
\begin{align}
    \int_0^1 dy\,\cos(\pi y)\ y^{3-\alpha} &= -\frac{\, _1F_2\left(2-\frac{\alpha }{2};\frac{1}{2},3-\frac{\alpha }{2};-\frac{\pi ^2}{4}\right)}{\alpha-4}\,, \quad \textrm{for} \quad \alpha<4\,,
\\
    \int_0^1 dy\,\cos(\pi y)\ (y+1)^{3-\alpha} &= -\frac{1}{2} \mathcal{K}_{\alpha -3}(1)+2^{3-\alpha } \mathcal{K}_{\alpha -3}(2)\,,
\\
    \int_0^1 dy\,\cos(\pi y)\ (y+2)^{3-\alpha} &= 2^{3-\alpha } \mathcal{K}_{\alpha -3}(2)-\frac{3^{4-\alpha }}{2} \mathcal{K}_{\alpha -3}(3)\,,
\\
    \int_0^1 dy\,\cos(\pi y)\ (y+3)^{3-\alpha} &= 2^{7-2 \alpha } \mathcal{K}_{\alpha -3}(4)-\frac{1}{2} 3^{4-\alpha } \mathcal{K}_{\alpha -3}(3)\,,
\end{align}\label{eq:int_solut_D4}
\end{subequations}
 with $_1F_2$ the generalized hypergeometric function and $\mathcal{K}_n(z)\equiv E_n(i\pi z) + E_n(-i\pi z)$, with $E_n(\pm i\pi z)$ denoting the exponential integral function, introduced to compactify notation. Combining Eqs.~\eqref{eq:AG_D4_asymp1}--\eqref{eq:AG_D4_asymp12} with the explicit solutions \eqref{eq:int_solut_D4} of the integrals, we obtain a simplified expression for the $(*)$ term~\eqref{eq:alg_connectivity_D4_intermediate_step2}, leading to the spectral gap scaling
\begin{align}\label{eq:algebraic_connectivity_4d}
    \delta \approx -\kappa_0 + \frac{8\zeta (\alpha -3)}{3}+\frac{16\zeta (\alpha -1)}{3} +  2^\alpha&\Bigg[\frac{3\ 2^{ -1} \mathcal{K}_{\alpha-3}(1)- 3^{4-\alpha } \mathcal{K}_{\alpha-3}(3)-2^{7-2\alpha }\mathcal{K}_{\alpha-3}(4)}{(\alpha -3) (\alpha -2) (\alpha -1)} \nonumber\\&\qquad-\frac{ \, _1F_2\left(2-\frac{\alpha }{2};\frac{1}{2},3-\frac{\alpha }{2};-\frac{\pi ^2}{4}\right)}{(\alpha -4)(\alpha -3) (\alpha -2) (\alpha -1)}\Bigg]n^{4-\alpha}.
\end{align}
Result~\eqref{eq:algebraic_connectivity_4d} accurately approximates the unscaled spectral gap $\delta$ when the number of lattice sites $n$ along each spatial dimension is sufficiently large, see Fig.~\ref{fig:asymptotic_d4}(b).\\

To normalize the spectral gap, we need to analyze the scaling of $\mathcal{E}(\vec{k}_{\max})$. Starting from the definition
\begin{equation}\label{eq:largest_ev_D4_step1}
    \mathcal{E}(\vec{k}_{\max}) =-\varepsilon_0 +\sum_{\vec{j} \neq\vec{0}}  \cos(\vec{k}_{\max}\cdot\vec{j})/\vert\vert \vec{j}\vert\vert_1^{\alpha}\ = -\varepsilon_0 + \hspace{-0.8cm}\sum_{\substack{j_1,j_2,j_3,j_4=-n/2+1;\\\vert j_1\vert + \vert j_2\vert+ \vert j_3\vert+ \vert j_4\vert\neq 0}}^{n/2} \hspace{-0.8cm} (\vert j_1\vert + \vert j_2\vert+ \vert j_3\vert+ \vert j_4\vert)^{-\alpha} \cos\left(\pi j_1 + \pi j_2 + \pi j_3+ \pi j_4\right),
\end{equation}
 the cosine may be replaced by $(-1)^{\ell}$, where $\ell = j_1+j_2+j_3+j_4$ for compactness, since $\ell\in \mathbb{Z}$. Then, upon expanding the quadruple summation and shifting the summation indices, we have
\begin{equation}\label{eq:largest_ev_D4_step2}
    \mathcal{E}(\vec{k}_{\max}) = -\varepsilon_0 +\hspace{-0.2cm}\underbrace{\sum_{\substack{j_1,j_2,j_3,j_4=0;\\\ell\neq 0}}^{n/2} \hspace{-0.6cm} \ell^{-\alpha} (-1)^{\ell}}_{(1)} + \underbrace{4 \sum_{j_1=1}^{n/2-1}\sum_{\substack{j_2,j_3,\\j_4=0}}^{n/2}\ell^{-\alpha} (-1)^{\ell}}_{(2)} + \underbrace{6 \sum_{j_1,j_2=1}^{n/2-1}\sum_{j_3,j_4=0}^{n/2} \ell^{-\alpha} (-1)^{\ell}}_{(3)}   + \underbrace{4\sum_{\substack{j_1,j_2,\\j_3=1}}^{n/2-1}\sum_{j_4=0}^{n/2} \ell^{-\alpha} (-1)^{\ell}}_{(4)}  + \underbrace{\sum_{\substack{j_1,j_2,\\j_3,j_4=1}}^{n/2-1} \ell^{-\alpha} (-1)^{\ell}}_{(5)}\,.
\end{equation}
The asymptotic scaling of the five terms, labeled $(1)$--$(5)$, can now be extracted by writing the quadruple summations as a series of single summations. In fact, in each expansion, only one term is non-negligible in the large-$n$ limit. Collectively representing terms with a negligible contribution by $\mathcal{C}_i$, with $i=1,2,3,4,5$ corresponding to term $(i)$ in Eq.~\eqref{eq:largest_ev_D4_step2}, we write the single summations that contribute significantly to $\mathcal{E}(\vec{k}_{\max})$ as
\begin{align}
    (1) &= \frac{1}{6}\sum_{x=1}^{n/2} (-1)^x \left(x^3+6 x^2+11 x+6\right) x^{-\alpha } + \mathcal{C}_1 \nonumber\\
    &\approx \frac{\left( 2^{3-\alpha } -2^{-1}\right) }{3}\zeta (\alpha -3) +(2^{3-\alpha } -1)\zeta (\alpha -2)+\frac{11\left( 2^{1-\alpha }-2^{-1} \right)}{3} \zeta (\alpha -1)+ \left(2^{1-\alpha } -1\right)\zeta (\alpha )\,,\nonumber
\\
    (2) &= \frac{2}{3} \sum_{x=1}^{n/2-1} (-1)^x (x+1)(x+2) x^{1-\alpha } + \mathcal{C}_2\approx \frac{1}{3} \left(2^{5-\alpha } -2\right)\zeta (\alpha -3) +\left(2^{4-\alpha } -2\right) \zeta (\alpha -2)+\frac{1}{3} \left(2^{4-\alpha } -2^2\right)\zeta (\alpha -1)\,,\nonumber
\\
    (3) &=  \sum_{x=1}^{n/2} (-1)^x (x-1)(x+1) x^{1-\alpha } + \mathcal{C}_3\approx \left(2^{4-\alpha } -1\right)\zeta (\alpha -3)-\left(2^{2-\alpha } -1\right)\zeta (\alpha -1)\,,\nonumber
\\
    (4) &= \frac{2}{3} \sum_{x=1}^{n/2+1} (-1)^x (x-1)(x-2) x^{1-\alpha } + \mathcal{C}_4
    \approx \frac{1}{3} \left(2^{5-\alpha } -2\right)\zeta (\alpha -3) -\left(2^{4-\alpha } -2\right) \zeta (\alpha -2)+\frac{1}{3} \left(2^{4-\alpha } -2^2\right)\zeta (\alpha -1)\,,\nonumber
\\\label{eq:nr5}
    (5) &= \frac{1}{6} \sum_{x=1}^{n/2+2} (-1)^x (x-3)(x-2)(x-1) x^{-\alpha } + \mathcal{C}_5\nonumber\\
    &\approx \frac{1}{3} \left(2^{3-\alpha }-2^{-1}\right) \zeta (\alpha -3) - \left(2^{3-\alpha } -1\right)\zeta (\alpha -2)+\frac{11}{3} \left(2^{1-\alpha } -2^{-1}\right)\zeta (\alpha -1) - \left(2^{1-\alpha } -1\right)\zeta (\alpha )\,.
\end{align}
Substituting the asymptotic results~\eqref{eq:nr5} into expression \eqref{eq:largest_ev_D4_step2}, we observe that, to leading-order in $n$, $\mathcal{E}(\vec{k}_{\max})$ scales as
\begin{equation}\label{eq:largest_eigenvalue_4d}
		\mathcal{E}(\vec{k}_{\max}) \approx -\varepsilon_0 + \frac{1}{3}\left(2^{7-\alpha}-2^3\right)\zeta(\alpha-3) +\frac{1}{3}\left(2^{6-\alpha}-2^4\right)\zeta(\alpha-1)\,.
\end{equation}
Figure \ref{fig:asymptotic_d4}(c) shows that this asymptotic result \eqref{eq:largest_eigenvalue_4d} provides a reliable description of the behavior of the Laplacian's largest eigenvalue, $\mathcal{E}(\vec{k}_{\max})$.

\appsection{Magnitude of the search fidelity}\label{app:fidelity}
Here we further discuss the asymptotic behavior of the order parameter $\chi_\alpha$ in the thermodynamic limit $N \to \infty$, which in turn determines the magnitude of the search fidelity, $F(T)=\vert\chi_\alpha\vert^2$. By definition, we have
\begin{equation}\label{eq:chi_def}
    \chi_\alpha = S_1^{(\alpha)}/\sqrt{S_2^{(\alpha)}}\,, \quad \mathrm{with} \quad S_\ell^{(\alpha)} = \frac{1}{N} \sum_{\vec{k} \neq \vec{0}} [\mathcal{E}_\alpha(\vec{k})]^{-\ell}\,,
\end{equation}
where $\mathcal{E}_\alpha(\vec{k})$ are the eigenvalues of the Laplacian $L_\alpha$. Noting that the hypercubic lattices are regular graphs with vertices of the same degree, we can set the energy shift $\varepsilon_0=0$ without loss of generality; see main text Eq.~(2). With this, the energies take the form $\mathcal{E}_\alpha(\vec{k}) = \sum_{\vec{j} \neq \vec{0}} \cos(\vec{k} \cdot \vec{j})/|\vec{j}|^\alpha$, where $|\vec{j}|$ denotes the Euclidean norm. To proceed, we consider $\chi_\alpha$ \eqref{eq:chi_def} for two regimes of the long-range tunneling exponent: (i) $0\leq\alpha<d$, see Sec.~\ref{a:ss:reg_i}, and (ii) $d<\alpha<3d/2$, see Sec.~\ref{a:ss:reg_ii}.

\subsection{Regime (i): $0\leq\alpha<d$}\label{a:ss:reg_i}
For strongly long-range tunneling, with exponent $\alpha\in[0,d)$, the dominant contribution to the summation $S_\ell^{(\alpha)}$ \eqref{eq:chi_def} comes from the eigenvalues $\mathcal{E}_\alpha(\vec{k})$ at low momentum values. Since $\vec{k}=\vec{0}$ is explicitly excluded, the main contribution will come from the second smallest eigenvalue, i.e.\ the spectral gap $\delta_\alpha$. The remaining eigenvalues scale with the lattice size $N$, hence the corresponding terms of $S_\ell^{(\alpha)}$, scaling as $1/\mathcal{E}_\alpha(\vec{k})$, tend to zero in the limit $N\rightarrow\infty$ for $\ell>0$. Consequently, we may write $S_\ell^{(\alpha)}\approx \frac{1}{2\pi}\int_\mathrm{BZ} \mathrm{d}\vec{k}\, \delta_\alpha^{-\ell} + c_\ell$, where the term $c_\ell$, containing contributions from the larger Laplacian eigenvalues, tends to zero for $N\rightarrow\infty$. It follows directly that $\chi_\alpha\approx \delta_\alpha^{-1}/\sqrt{\delta_\alpha^{-2}}=1$.

\subsection{Regime (ii): $d<\alpha<3d/2$}\label{a:ss:reg_ii}
 The behavior of $\chi_\alpha$ \eqref{eq:chi_def} strongly depends on the momentum-dependence of the Laplacian eigenvalues $\mathcal{E}_\alpha(\vec{k})$ around $\vec{k}=\vec{0}$, which we extract by approximating the summations by integrals, $\sum_{\vec{j}}\rightarrow\int \mathrm{d}\vec{j}$, with the integration extending over the entire space, excluding the origin $\vec{j} = \vec{0}$. Considering each lattice dimension $d\leq4$ independently, we perform an appropriate coorindate transformation and derive the following leading-order results:
\begin{equation}\label{eq:E_k_dep_generic}
    \mathcal{E}_\alpha(\vec{k}) \propto \begin{cases}
				\vert \vec{k}\vert^{\alpha -d}\,, \quad & \alpha\in(d,d+2)\\
				\vert\vec{k}\vert^2\,, & \alpha > d+2
			\end{cases}\,.
\end{equation}
Detailed calculations are provided in the itemized list below.
	\begin{enumerate}
		\item In one dimension $\mathcal{E}_\alpha(k)\approx \int_1^\infty \mathrm{d}j\, j^{-\alpha} \cos{kj}$.  For $\alpha>0$ and $k\in\mathbb{R}$ the integral evaluates to
		\begin{equation}
			\int_1^\infty \mathrm{d}j\, j^{-\alpha} \cos{kj} = \frac{\, _1F_2\left(\frac{1}{2}-\frac{\alpha }{2};\frac{1}{2},\frac{3}{2}-\frac{\alpha }{2};-\frac{k^2}{4}\right)}{\alpha -1}+\sin \left(\frac{\pi  \alpha }{2}\right) \Gamma (1-\alpha ) | k| ^{\alpha -1}\,,
		\end{equation}
		with $_pF_q(a;b;z)$ the generalized hypergeometric function and $\Gamma$ the $\Gamma$ function. Expanding in powers of the momentum $k$, we find
		the leading-order behavior of the dispersion relation
		\begin{equation}\label{eq:Ealpha_approx_D1}
			\mathcal{E}_\alpha(k) \propto \begin{cases}
				\sin \left(\frac{\pi  \alpha }{2}\right) \Gamma (1-\alpha ) \vert k\vert^{\alpha -1}\,, \quad & \alpha\in(1,3)\\
				-\frac{1}{2 (\alpha -3)} k^2\,, & \alpha >3
			\end{cases}.
		\end{equation}
		
		\item For two spatial dimensions, the eigenenergies can be approximated as $\mathcal{E}_\alpha(\vec{k})\approx\int_0^\infty \mathrm{d}j_1\, \int_0^\infty \mathrm{d}j_2\, (j_1^2 + j_2^2)^{-\alpha/2} \cos(j_1 k_1 + j_2 k_2)$, where we require $\vert\vec{j}\vert \neq 0$. The latter requirement is included explicitly later. Converting to polar coordinates, the area element transforms as $ \mathrm{d}j_1  \mathrm{d}j_2 = r  \mathrm{d}r  \mathrm{d}\theta$ with $r = \vert \vec{j}\vert = \sqrt{j_1^2+j_2^2}$ and $\theta = \arctan(j_2/j_1)$, and we find the integral
		$\int_1^\infty\mathrm{d}r\, r^{1-\alpha}\int_0^{\pi}\mathrm{d}\theta\,\cos(r k \cos(\theta-\theta_k))$ where $k\equiv \vert\vec{k}\vert$ and $\theta_k=\arctan(k_2/k_1)$ is the angle providing the direction of the momentum vector $\vec{k}$. To evaluate the angular part of the integral, we perform the change of variable $\theta  \rightarrow \theta' + \theta_k$, yielding the simplified form $\int_1^\infty\mathrm{d}r\, r^{1-\alpha}\int_0^{\pi}\mathrm{d}\theta'\,\cos(r k \cos(\theta'))$. The cosine term $\cos(r k \cos(\theta'))$ can be rewritten using the Jacobi–Anger expansion. In its most generic form, $e^{iz\cos(\theta)}\equiv \sum_{n=-\infty}^{+\infty} i^n J_n(z) e^{in\theta}$ with $J_n(z)$ the $n$-th Bessel function of the first kind, giving a convenient expansion of exponentials of trigonometric functions in the basis of their harmonics. We use the real-valued variation $\cos(z\cos(\theta))\equiv J_0(z)+2\sum_{n=1}^{\infty}(-1)^nJ_{2n}(z)\cos(2n\theta)$. The angular integral is now easily evaluated, leaving only the integral over the radial part: 
		\begin{equation}
			\pi\int_1^\infty\mathrm{d}r\, r^{1-\alpha} J_0(rk) = \pi \left(\frac{\, _1F_2\left(1-\frac{\alpha }{2};1,2-\frac{\alpha }{2};-\frac{k^2}{4}\right)}{\alpha -2}-\frac{2^{-\alpha}  \alpha\,  \Gamma \left(-\frac{\alpha }{2}\right) \vert k\vert^{\alpha -2}}{\Gamma \left(\frac{\alpha }{2}\right)}\right)
		\end{equation}
		for $k>0$, $k\in\mathbb{R}$ and $\alpha>1/2$. Performing a series expansion around $k=0$, we obtain, to leading order in $k$,
		\begin{equation}\label{eq:Ealpha_approx_D2}
			\mathcal{E}_\alpha(\vec{k}) \propto \begin{cases}
				-\frac{ 2^{-\alpha -1} \pi\alpha  \Gamma \left(-\alpha /2\right)}{\Gamma \left(\alpha/2\right)} \vert k\vert^{\alpha -2}\,, \quad & \alpha\in(2,4)\\
				-\frac{\pi}{8 (\alpha -4)} k^2\,, & \alpha > 4
			\end{cases}.
		\end{equation}
		
		\item For $d=3$ a similar procedure may be followed. After approximating the summation over $\vec{j}$ by an integral, we convert to spherical coordinates, with the volume element expressed as $\mathrm{d}\vec{j} = r^2 \sin\theta \, \mathrm{d}r \, \mathrm{d}\theta \, \mathrm{d}\phi$, $r=\vert\vec{j}\vert$, such that $\mathcal{E}_\alpha(\vec{k})\approx \int_1^\infty \mathrm{d}r\, r^{2-\alpha} \int_0^\pi \mathrm{d}\theta\,\cos(kr \cos\theta) \sin\theta   \int_0^{2\pi} \mathrm{d}\phi$. Due to the system being isotropic, we assumed $\vec{k}$ is aligned along the $z$-axis, allowing for simplification of the angular integrals. The azimuthal integral evaluates to $2\pi$, while the angular integral is $\int_0^\pi \mathrm{d}\theta\,\cos(kr \cos\theta) \sin\theta   = \sin(kr)/kr$. Substituting this result, we compute the radial component as
		\begin{equation}
			\int_1^\infty \mathrm{d}r\, \frac{\sin(kr)}{kr} r^{2-\alpha} = \frac{\, _1F_2\left(\frac{3}{2}-\frac{\alpha }{2};\frac{3}{2},\frac{5}{2}-\frac{\alpha }{2};-\frac{k^2}{4}\right)}{\alpha -3}+\sin \left(\frac{\pi  \alpha }{2}\right) \Gamma (2-\alpha ) | k| ^{\alpha -3}
		\end{equation}
		for $k\in\mathbb{R}$ and $\alpha>1$. A series expansion then yields the leading-order behavior of the Laplacian eigenenergies:
		\begin{equation}\label{eq:Ealpha_approx_D3}
			\mathcal{E}_\alpha(\vec{k}) \propto \begin{cases}
				2 \pi  \sin \left(\frac{\pi  \alpha }{2}\right) \Gamma (2-\alpha ) \vert k\vert^{\alpha -3}\,, \quad & \alpha\in(3,5)\\
				-\frac{\pi  }{3 (\alpha -5)} k^2\,, & \alpha > 5
			\end{cases}.
		\end{equation}
	
		\item In four spatial dimensions, we approximate the eigenenergy summation by an integral and then transform to hyper-spherical coordinates, with the volume element $d\vec{j} = r^3 \sin^2\theta \, dr \, d\theta \, d\phi_1 \, d\phi_2$. After simplification we obtain $\mathcal{E}_\alpha(\vec{k}) \approx 4\pi \int_1^\infty \mathrm{d}r\,r^{3-\alpha} \int_0^\pi \mathrm{d}\theta\,\cos(kr \cos\theta) \sin^2\theta$. The remaining angular integral can be evaluated explicitly, giving $\int_0^\pi\mathrm{d}\theta\, \cos(kr \cos\theta) \sin^2\theta = \pi  J_1(k r)/(k r)$ with $J_n(z)$ the Bessel function of the first kind. Inserting this result into the radial integral leads to
        \begin{equation}
           4\pi^2 \int_1^\infty \mathrm{d}r\,\frac{  J_1(k r)}{k r} r^{3-\alpha } = \pi ^2 \Gamma \left(2-\frac{\alpha }{2}\right) \left(\frac{2^{4-\alpha } \vert k\vert^{\alpha -4}}{\Gamma \left(\frac{\alpha }{2}\right)}-\, _1\tilde{F}_2\left(2-\frac{\alpha }{2};2,3-\frac{\alpha }{2};-\frac{k^2}{4}\right)\right)
        \end{equation}
        for $k>0$, $k\in\mathbb{R}$ and $\alpha>1.5$, and where $_1\tilde{F}_2$ is the regularized generalized hypergeometric function and $\Gamma$ is the $\Gamma$ function, as before. The leading-order behavior of the eigenvalues is then extracted as
        \begin{equation}\label{eq:Ealpha_approx_D4}
			\mathcal{E}_\alpha(\vec{k}) \propto \begin{cases}
				\frac{\pi ^2 2^{4-\alpha } \Gamma \left(2-\frac{\alpha }{2}\right) }{\Gamma \left(\frac{\alpha }{2}\right)} \vert k\vert^{\alpha -4}\,, \quad & \alpha\in(4,6)\\
				-\frac{\pi ^2 }{4(\alpha -6)} k^2\,, & \alpha > 6
			\end{cases}.
		\end{equation}
    \end{enumerate}
Combining the results for all spatial dimensions $d\in[1,4]$, we obtain Eq.~\eqref{eq:E_k_dep_generic}. 

The next step involves computing $S_\ell^{(a)}$. Since we are working in the thermodynamic limit, we approximate the discrete momentum space by a continuum:
\begin{equation}\label{eq:S_l_int}
    S_\ell^{(\alpha)} \approx \frac{1}{2\pi} \int_{\mathrm{BZ}} \mathrm{d}\vec{k}\, \vert\vec{k}\vert^{-(\alpha-d)\ell},\quad \alpha\in(d,d+2)\,.
\end{equation}
 The prefactors coming from the eigenenergy approximation, Eqs.~\eqref{eq:Ealpha_approx_D1}, \eqref{eq:Ealpha_approx_D2}, \eqref{eq:Ealpha_approx_D3} and \eqref{eq:Ealpha_approx_D4} and summarized in Eq.~\eqref{eq:E_k_dep_generic}, do not contribute to the ratio of interest \mbox{$\chi_\alpha = S_1^{(\alpha)}/\sqrt{S_2^{(\alpha)}}$} and are therefore neglected. Notice now that the integrand \eqref{eq:S_l_int} depends solely on the magnitude of the momentum vector $\vec{k}$, allowing for the transformation to spherical coordinates in $d$ dimensions, $\vert \vec{k}\vert=r$ and $\mathrm{d}\vec{k}=\Omega_d r^{d-1} \mathrm{d}r$, with $\Omega_d$ the surface area of the unit sphere in $d$ dimensions. Collecting the global prefactors and denoting it by $c_d$, we evaluate the integral:
 \begin{equation}\label{eq:S_l_ana}
     S_\ell^{(\alpha)} \approx\frac{c_d}{d+\ell(d-\alpha)}\,.
 \end{equation}
The convergence criterion is $\mathrm{Re}[d+\ell(d-\alpha)]>0$. For the cases we consider, the largest value of $\ell$ is $\ell = 2$. This implies that the integral $S_2^{(\alpha)}$ diverges for $\alpha > 3d/2$. The value of $\chi_\alpha= S_1^{(\alpha)}/\sqrt{S_2^{(\alpha)}}$ in the regime $\alpha > 3d/2$ is therefore determined by the rate of divergence of $\sqrt{S_2^{(\alpha)}}$ compared to $S_1^{(\alpha)}$. More precisely, we find that $\chi_\alpha \to 0$. Instead, in the regime $\alpha\in(d,3d/2)$, Eq.~\eqref{eq:S_l_ana} leads to
\begin{equation}
    \chi_\alpha = \frac{S_1^{(\alpha)}}{\sqrt{S_2^{(\alpha)}}} \approx \sqrt{c_d} \frac{\sqrt{3-2\alpha/d}}{2-\alpha/d}\,, 
\end{equation}
exhibiting the behavior of Fig.~3(d) of the main text, where $\chi_\alpha$ acts as an order parameter and decreases monotonically to zero for $d<\alpha<3d/2$. The coefficient $c_d$ can be determined numerically. 

\twocolumngrid


\providecommand{\noopsort}[1]{} \providecommand{\noopsort}[1]{}

\end{document}